\documentclass[letterpaper,twocolumn,10pt]{article}
\usepackage{usenix}

\usepackage{graphicx}
\usepackage{subfigure}
\usepackage{amsmath}
\usepackage{amsthm}
\usepackage{amssymb}
\usepackage{mathrsfs}
\usepackage{slashbox}
\usepackage{float}
\usepackage[vlined,ruled,linesnumbered]{algorithm2e}
\usepackage{xcolor}
\usepackage{multicol}
\usepackage{multirow}

\newtheorem{thm}{Theorem}

\begin{document}
%
\title{Label-only Model Inversion Attack: The Attack that Requires the Least Information}


\author{
{\rm Dayong Ye}\\
UTS, Australia\\
Dayong.Ye@uts.edu.au
\and
{\rm Tianqing Zhu}\\
UTS, Australia\\
Tianqing.Zhu@uts.edu.au
\and
{\rm Shuai Zhou}\\
UTS, Australia\\
12086892@uts.edu.au
\and
{\rm Bo Liu}\\
UTS, Australia\\
Bo.Liu@uts.edu.au
\and
{\rm Wanlei Zhou}\\
City University of Macau, China\\
wlzhou@cityu.mo
\and
} 

\maketitle

\begin{abstract}
In a model inversion attack, an adversary attempts to reconstruct the data records, used to train a target model, using only the model's output. 
In launching a contemporary model inversion attack, the strategies discussed are generally based on 
either predicted confidence score vectors, i.e., black-box attacks, or the parameters of a target model, i.e., white-box attacks. 
However, in the real world, model owners usually only give out the predicted labels; 
the confidence score vectors and model parameters are hidden as a defense mechanism to prevent such attacks. 
Unfortunately, we have found a model inversion method that can reconstruct the input data records based only on the output labels. 
We believe this is the attack that requires the least information to succeed and, therefore, has the best applicability. 
The key idea is to exploit the error rate of the target model to compute the median distance 
from a set of data records to the decision boundary of the target model. 
The distance, then, is used to generate confidence score vectors 
which are adopted to train an attack model to reconstruct the data records. 
The experimental results show that highly recognizable data records can be reconstructed with far less information than existing methods.
\end{abstract}

\section{Introduction}

Machine learning (ML) models, like deep neural networks (DNN), have been applied to a wide range of applications 
from computer vision to speech recognition \cite{Salem19}. 
However, building a powerful ML model, requires a massive amount of training data, and these data may contain sensitive information 
that is vulnerable to a variety of adversarial activities \cite{Tramer16}. 
One common type of attack is a model inversion attack \cite{Fred14,Fred15}, 
where the aim is to reconstruct some of the model's training data based on its outputs or parameters \cite{Yang19,Zhang20}. 

Existing model inversion attacks can be roughly classified into two types 
based on which of this information they exploited: black-box attacks exploit outputs and white-box attacks exploit parameters. 
In black-box attacks, an attacker exploits only the predictions of a target ML model 
but cannot explore the structure and parameters of the model \cite{Yang19}. 
By comparison, in white-box attacks, an attacker can utilize the full information of a target model, 
notably its architecture, parameters, and gradients \cite{Nasr19,Geiping20}. 
The performance of white-box attacks is better than black-box attacks, 
since white-box attackers have access to and use more information than black-box attackers. 
However, white-box attacks may have limited applicability 
when there is no access to the internal information of a target model, such as Amazon Rekognition API \cite{Amazon}. 
Our concern is with the black-box model inversion attacks, as these can be launched far more widely.

For a black-box attacker, the only available information is 
the label and confidence score vector predicted by a target model 
when classifying a given data record. 
The label indicates the class to which the given data record belongs, 
while the confidence score vector is a probability distribution over the available classes. 
Each score indicates the confidence in predicting the corresponding class. 
In a typical black-box model inversion attack, 
an adversary trains an attack model to accurately reconstruct an input data record using only the confidence score vector. 
However, in extreme cases, even confidence score vectors are hidden by the owner of target models, 
who instead publishes only the predicted labels of the input data records \cite{Choo20}. 
Recently, label-only membership inference attacks have been studied \cite{Choo20,Li21}
which aim to infer whether a given data record is in the training data of a target model. 
However, to the best of our knowledge, no label-only model inversion attack has actually been conducted. 
By comparison with label-only membership inference attacks, 
performing a label-only model inversion attack is much more challenging for several reasons. 
First, reconstructing a data record is much more difficult than simply inferring whether the record is or is not in a model's training \cite{Yeom20}; 
and second, predicted labels contain so little information that reconstructing a data record from a label is much harder 
than reconstructing it from a confidence score vector \cite{Wu16}. 
For example, an attacker trying to infer membership from only a label 
can use the distance between a given data record and the target model's decision boundary 
to directly infer whether this record is in the target model's training set. 
If the distance is larger than a pre-defined threshold, this record is deemed to be in, and lower than the threshold - it is out. 
However, for a label-only model inversion attacker, using only this distance is insufficient to reconstruct the given record. 

In this paper, we propose a novel label-only model inversion attack method 
that can reconstruct data records based only on their labels predicted by a target model, 
even if the attacker has never seen the data records. 
To reconstruct data records, we first train shadow models to recover confidence score vectors 
and then train an attack model that takes the confidence vectors as input and outputs reconstructed data records. 
In summary, we make the following four contributions. 
 First, we are the first to study label-only model inversion attacks 
    where the attacker has the least information compared with existing attack methods. 
    Thus, our research shows that even if an ML model hides everything but the labels, their training data are still vulnerable.
    Second, we associate label-only model inversion attacks with the error rate of models, 
    and develop a vector recovery approach by using the error rate of a target model.
    Third, we conduct comprehensive experiments, which show that our method works very well 
    even if the attacker can only observe the output labels of the target model and 
    the attacker has never seen the training data of the target model.
    Last, we also propose a defense method against our attack 
    and show that our attack method can still achieve reasonable performance. 

\begin{figure}[ht]
\centering
	\includegraphics[scale=0.3]{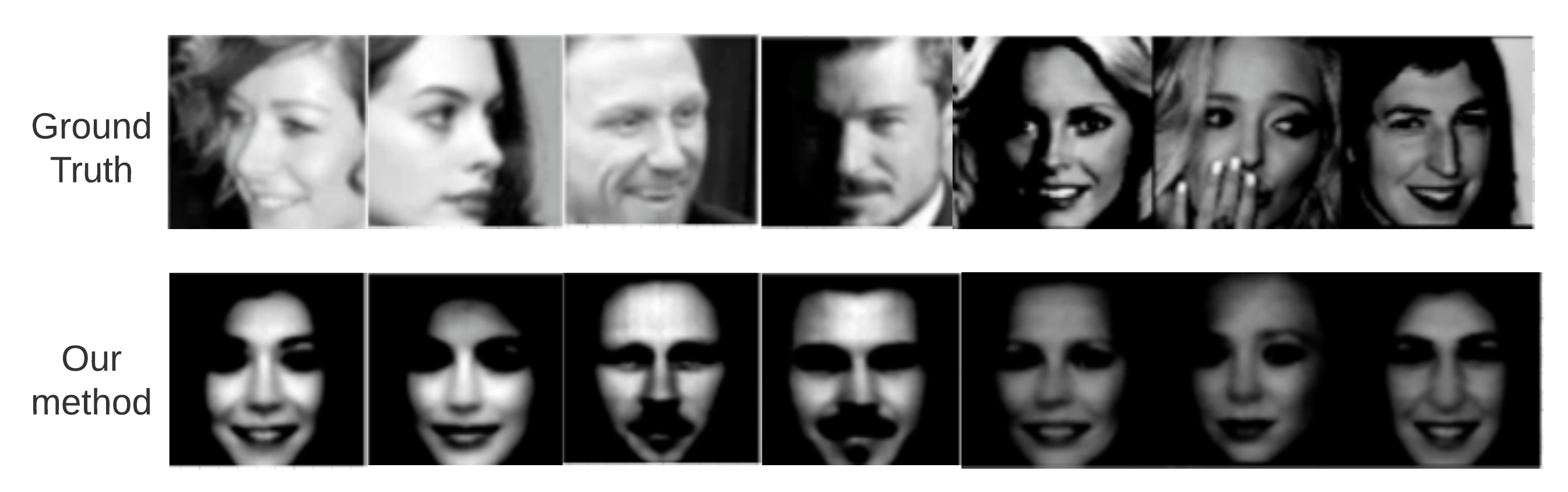}
	\caption{The first row is the ground truth and the second row is the reconstructed images. The results show that even if the classifier outputs only labels and the attacker can collect only low-quality auxiliary samples, e.g., profiles, a representative facial image for each person can still be reconstructed.}
	\label{fig:example}
\end{figure}


\begin{figure*}[ht]
\centering
	\includegraphics[scale=0.42]{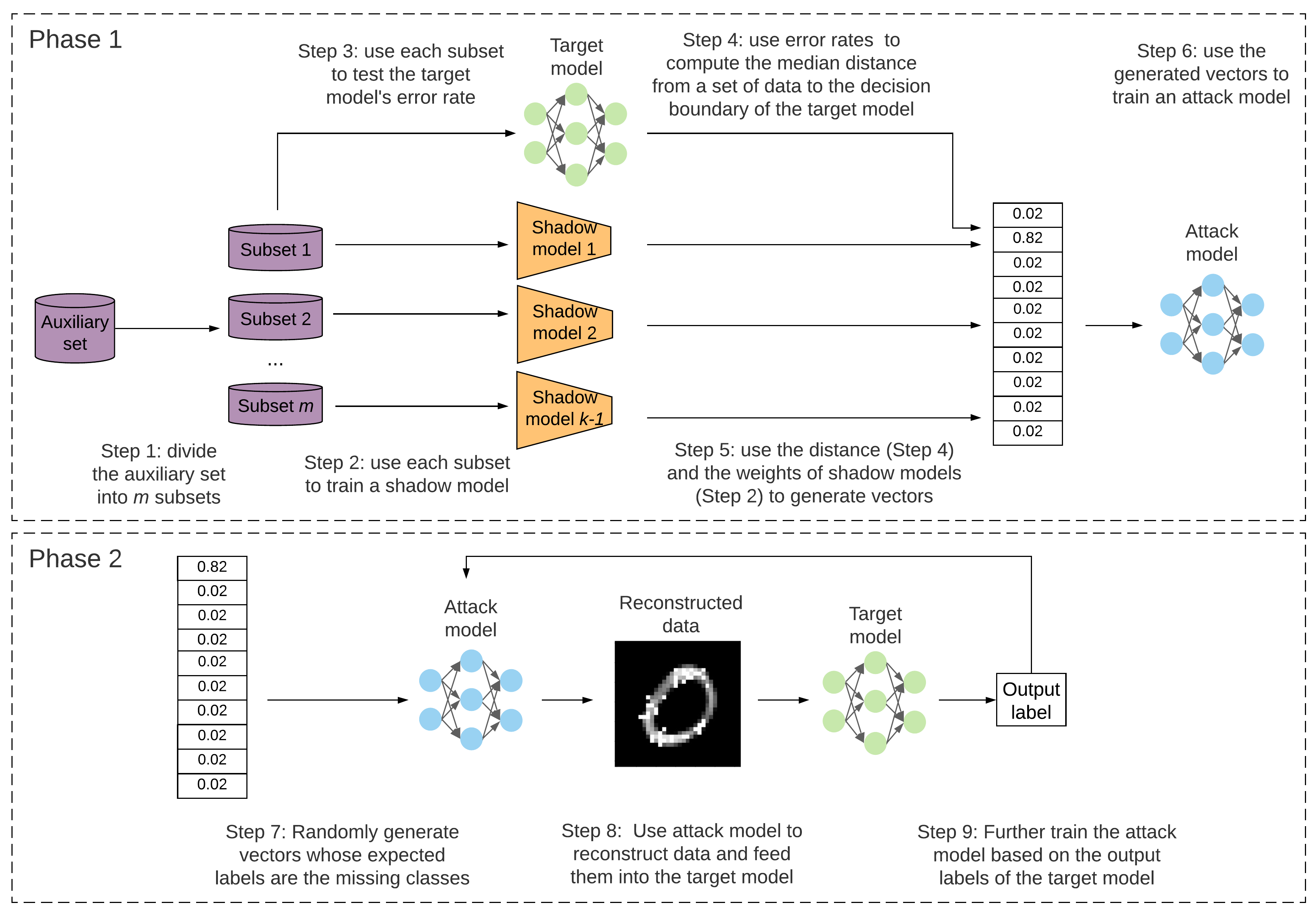}
	\caption{Overview of our \emph{label-only} attack method. Once an attacker receives the label of the input data record $\mathbf{x}^*$, he collects an auxiliary set $D_{aux}$ of samples which belong to the same class as the input record $\mathbf{x}^*$. The attacker uses $D_{aux}$ with Gaussian noise to access the target model $T$ to test its error rate $\mu$ (Step 1) which is used to compute the distance $d$ between $\mathbf{x}^*$ and the decision boundary of $T$ (Step 2). The attacker then uses $D_{aux}$ to train a shadow model (Step 3) whose weights, associated with distance $d$, are exploited to approximately recover the confidence score vector (Step 4). Finally, the recovered vector is adopted to train an attack model $A$ to reconstruct $\mathbf{x}^*$ (Step 5).}

	\label{fig:overview}
\end{figure*}

\section{Preliminary}
\subsection{Model inversion attacks}
Model inversion attacks aim to reconstruct a target model's input data from its output \cite{Fred15}.
Typically, the attacker trains a new attack model on an auxiliary dataset, 
which acts as the inverse of the target model \cite{Yang19}. 
The attack model takes the output of the target model as input 
and tries to reconstruct the original input data.

Formally, let $T$ be the target model consisting of $m$ classes, and $A$ be the attack model. 
Given a data record $(\mathbf{x},y)$, where $\mathbf{x}$ is the data and $y$ is the label of $\mathbf{x}$, 
the attacker inputs $\mathbf{x}$ into $T$ and receives $T(\mathbf{x})$. 
$T(\mathbf{x})$ is then fed into $A$ which returns $A(T(\mathbf{x}))$ that is expected to be very similar to $\mathbf{x}$. 
In other words, the aim is to minimize the following objective: 
\begin{equation}
    C(A)=\mathbb{E}_{\mathbf{x}\sim p_{\mathbf{x}}}[L(A(T(\mathbf{x})),\mathbf{x})],
\end{equation}
where $p_{\mathbf{x}}$ is the underlying probability distribution of $\mathbf{x}$, and $L$ is a loss function such as an $L_2$ loss.

\subsection{Threat model}
Consider now an attacker that can only observe a target model's output labels. 
That would mean the attacker could only reconstruct a semantically meaningful data record for each class of the target model. 
This type of attack is also known as the class inference attack \cite{Fred15,Yang19}. 
For any given data record $(\mathbf{x},y)$, the attacker can only observe the target model's output label, $T(\mathbf{x})$. 
The attacker does not have any data from the training set of the target model, $D^{train}_{target}$. 
All they have is an auxiliary dataset that shares the same distribution as the training set of the target model: 
$D_{aux}\sim_{p_{\mathbf{x}}}D^{train}_{target}$ and $D_{aux}\cap D^{train}_{target}=\emptyset$. 
This assumption is commonly made in the literature related to this type of attack \cite{Shokri17,Salem19,Yang19,Li21}. 
The attacker aims to reconstruct a recognizable and representative sample for each class of the target model.

However, we have relaxed this assumption
by enabling the attacker to collect only low-quality data records, 
as we believe this more closely imitates real-world situations. 
For instance, in our motivating example, the attacker has access to facial images but only profiles to represent the low quality. 
The model is a facial recognition classifier, and the attacker is curious as to what the people, in the training set of the classifier, look like. 
The attacker, however, can collect only profiles of these people from the Internet. 
Further, the classifier outputs only labels, which thwarts most contemporary attack methods. 
However, we will demonstrate that, with our attack method, 
the attacker can reconstruct representative facial images of these people as shown in Figure \ref{fig:example}. 

Note that although we have used facial recognition as our example, 
this research does not focus on face restoration or deblurring \cite{Dahl17,Shen18,Menon20}. 
Those works focus on recovering a noisy or low-resolution face from a given photo, 
while our premise is to investigate methods of inferring the general features of classes given a target model. 
In this way, our research has the potential to be applied beyond image classifiers to non-image classifiers as well. 
For example, an organization uses its private dataset, e.g., its subscribers' movie viewing histories, 
to train a classifier, e.g., a movie recommendation system. 
The organization then releases this dataset but hiding sensitive information, e.g., its subscribers' names and addresses \cite{Ghoshal20}. 
Given that the classifier outputs only labels, with our method, an attacker, using this released dataset as the auxiliary set, 
may still be able to recover the sensitive information. 
We leave the research of inverting non-image classifiers as our future work. 


\section{Attack method design}\label{sec:method}
\subsection{Overview of our method} 
Figure \ref{fig:overview} presents an overview of our attack method which consists of the following five steps. 

\textbf{Step 1.} For a given target data record $\mathbf{x}^*$, the attacker collects a set of data records $D_{aux}$
that belong to the same class as $\mathbf{x}^*$. 
The attacker then adds Gaussian noise to each record in $D_{aux}$, 
and feeds the noisy set to the target model $T$ to test $T$'s classification error rate $\mu$.

\textbf{Step 2.} The attacker uses $\mu$ to compute the distance $d$ from $\mathbf{x}^*$     to the decision boundary of model $T$. 
    
\textbf{Step 3.} The attacker adopts the dataset $D_{aux}$ to train a shadow model $S$. 
    
\textbf{Step 4.} The attacker uses the weights of model $S$ and the distance $d$ 
    to approximately compute the confidence score vector $\mathbf{y}$ of record $\mathbf{x}^*$. 
    
\textbf{Step 5.} Finally, the attacker uses $D_{aux}$ to train an attack model $A$ 
    to reconstruct the target record $\mathbf{x}^*$. 

\subsection{Step 1: test the target model $T$'s classification error rate}
\begin{algorithm}
\caption{Test the target model $T$'s error rate}
\label{alg:step1}
\textbf{Input}: The label of record $\mathbf{x}^*$;\\
\textbf{Output}: The classification error rate of target model $T$;\\
Initialize $\hat{D}_{aux}=\emptyset$;\\
Collect a set of records $D_{aux}$ which have the same class as $\mathbf{x}^*$;\\
\For{each $\mathbf{x}\in D_{aux}$}{
    $\hat{\mathbf{x}}\leftarrow\mathbf{x}+\mathcal{N}(0,\sigma^2\mathbf{I})$;\\
    $\hat{D}_{aux}\leftarrow\hat{D}_{aux}\cup\{\hat{\mathbf{x}}\}$;\\
}
Feed $\hat{D}_{aux}$ to target model $T$ and receive the error rate $\mu$;\\
\end{algorithm}
The first step of our method is formalized in Algorithm \ref{alg:step1}. 
Given that the attacker receives the output label of target record $\mathbf{x}^*$, 
a set of records $D_{aux}$ can be collected that are in the same class as $\mathbf{x}^*$ (Line 4). 
The attacker then adds Gaussian noise to each record $\mathbf{x}$ in $D_{aux}$, 
and forms a new set $\hat{D}_{aux}$ (Lines 5-7). 
The attacker feeds $\hat{D}_{aux}$ into model $T$ to test its error rate $\mu$ (Line 8). 
The error rate $\mu$ will be used to compute the distance from 
$\mathbf{x}^*$ to the decision boundary of model $T$ in Step 2.

The key component in Algorithm \ref{alg:step1} is Line 6, 
$\hat{\mathbf{x}}\leftarrow\mathbf{x}+\mathcal{N}(0,\sigma^2\mathbf{I})$, 
adding Gaussian noise to each record in $D_{aux}$. 
Theorem \ref{thm:gaussian} highlights that a data record's distance to the boundary 
is directly related to the model's error rate when the point is perturbed 
by isotropic Gaussian noise \cite{Gilmer19,Choo20}. 
Therefore, to compute the distance, the data records perturbed with Gaussian noise are used to test target model $T$'s error rate. 
More discussion on the use of Gaussian noise will be given in the next section.
\begin{thm}[Gaussian Isoperimetric Inequality \cite{Gilmer19}]\label{thm:gaussian}
Let $q=\mathcal{N}(0,\sigma^2\mathbf{I})$ be the Gaussian distribution 
on $\mathbb{R}^n$ with a variance of $\sigma^2\mathbf{I}$, and for some error set $E\subseteq\mathbb{R}^n$, 
let $\mu=\mathbb{P}_{\mathbf{x}\sim q}[\mathbf{x}\in E]$. 

Then, let $\Phi(t)=\frac{1}{\sqrt{2\pi}}\int_{-\infty}^t exp(\frac{-x^2}{2})dx$ be the cumulative distribution function of the 
standard normal distribution, and let $d^*_q(E)$ be the median distance 
from one of the noisy records to the nearest error, i.e., the boundary. 
If the error rate $\mu\geq\frac{1}{2}$, then $d^*_q(E)=0$. 
Otherwise, $d^*_q(E)\leq-\sigma\Phi^{-1}(\mu)$ with equality when $E$ is a half space.
\end{thm}

\subsection{Step 2: compute the distance $d$ from $\mathbf{x}^*$ to $T$'s decision boundary}
After Step 1, we have the classification error rate $\mu$ of the target model $T$. 
According to Theorem \ref{thm:gaussian}, the distance $d$ from the record $\mathbf{x}^*$ to the decision boundary of the model $T$ can be computed as 
\begin{equation}\label{eq:distance1}
    d=-\sigma\Phi^{-1}(\mu), 
\end{equation}
where $\Phi^{-1}(\mu)<0$ when $\mu<1$. 
This distance interprets the confidence of the target model $T$ in classifying the record $\mathbf{x}^*$. 
If this distance is large, i.e., far away from the decision boundary, 
it means model $T$ is very confident in classifying $\mathbf{x}^*$. 
Therefore, the corresponding confidence score of the output label of $\mathbf{x}^*$ will be high. 
A larger distance implies a higher confidence score. 
To compute the confidence score, the attacker needs to train a shadow model $S$.

\subsection{Step 3: train a shadow model $S$}
The attacker trains a shadow model $S$ using dataset $D_{aux}$. 
Model $S$ is a binary linear regression model that outputs a confidence score $h(\mathbf{x})$ for an input $\mathbf{x}$ in the form of: 
\begin{equation}
    h(\mathbf{x})=\alpha(\mathbf{w}^{\top}\mathbf{x}+b),
\end{equation}
where $\alpha(t)=\frac{1}{1+e^{-t}}\in(0,1)$ is the logistic function. 
To train the shadow model $S$, the samples in $D_{aux}$ are labeled as ``$1$''. 
Moreover, the attacker also randomly collects some extra data as negative samples, $D_{neg}$, 
that do not belong to the same class as $\mathbf{x}^*$. 
Those negative samples are labelled as ``$0$'' and will be discarded after training the shadow model. 
These labels are used as ground truth, and the binary cross-entropy loss function is used to train shadow model $S$. 

For a linear model, there is a monotone relationship between 
the model's confidence in a data record $\mathbf{x}$ and 
the Euclidean distance from $\mathbf{x}$ to the model's decision boundary. 
As analyzed in \cite{Chen20,Choo20}, the distance from $\mathbf{x}$ to the boundary can be computed as 
\begin{equation}\label{eq:distance2}
    d=\frac{\mathbf{w}^{\top}\mathbf{x}+b}{||\mathbf{w}||_2}=\frac{\alpha^{-1}(h(\mathbf{x}))}{||\mathbf{w}||_2}.
\end{equation}
Therefore, for a linear model, obtaining a data record's distance 
to the boundary yields the same information as the model's confidence score in classifying that record. 
The attacker is interested in the confidence score, $h(\mathbf{x}^*)$. 
However, the attacker does not have an $\mathbf{x}^*$ that is one of the training samples of the target model $T$. 
Thus, $\mathbf{x}^*$ cannot be fed into shadow model $S$ to directly receive $h(\mathbf{x}^*)$. 
Nevertheless, the value of $h(x^*)$ can be approximately computed using the weights of the shadow model 
and the error rate of the target model classifying the samples in $D_aux$.

As claimed in \cite{Gilmer19,Song19}, deep and non-linear models can be closely approximated by linear models in the vicinity of the data. 
Formally, the following theorem supports this claim. 
\begin{thm}[\cite{Zhang17}]\label{thm:approximatity}
A neural network $C$ can represent any function of a dataset $D$ of size $z$ in $d$ dimensions, 
if for every dataset $D\subseteq\mathbb{R}^d$ with $|D|=n$ and every function $f:D\rightarrow\mathbb{R}$, 
there exists a setting for the weights of $C$ such that $C(\mathbf{x})=f(\mathbf{x})$ for every $\mathbf{x}\in D$.
\end{thm}
Theorem \ref{thm:approximatity} can be explained inversely: 
if a function $f$ can be found such that $f(\mathbf{x})=C(\mathbf{x})$ for every $\mathbf{x}\in D$, 
then the neural network $C$ can be represented by the function $f$ on dataset $D$. 
In our problem, the target model $T$ can be interpreted as a neural network $C$ and the shadow model $S$ can be interpreted as the function $f$. 
Also, the auxiliary set $D_{aux}$ can be interpreted as a given dataset $D$. 
Hence, the shadow model $S$ can represent the target model $T$ of the auxiliary set $D_{aux}$, 
if $S(\mathbf{x})=T(\mathbf{x})$ for every $\mathbf{x}\in D_{aux}$. 
Based on this analysis, the distance computed in Eq. \ref{eq:distance2} can be used to 
approximate the distance between $\mathbf{x}^*$ and the decision boundary of model $T$. 
Then, by equating Eqs. \ref{eq:distance1} and \ref{eq:distance2}, we have: 
\begin{equation}
    d=-\sigma\Phi^{-1}(\mu)=\frac{\alpha^{-1}(h(\mathbf{x}^*))}{||\mathbf{w}||_2}, 
\end{equation}
which is used to compute the confidence score $h(\mathbf{x}^*)$ as: 
\begin{equation}\label{eq:confidence}
    h(\mathbf{x}^*)=\frac{1}{1+exp[\sigma\Phi^{-1}(\mu)||\mathbf{w}||_2]}.
\end{equation}
A detailed deduction of Eq. \ref{eq:confidence} is given in the next section.

\subsection{Step 4: recover the confidence score vector $\mathbf{y}$}
As the input of the attack model $A$ is a confidence score vector rather than an individual confidence score, 
the attacker must recover the confidence score vector from a confidence score obtained in Step 3. 
Since the attacker only has information about the label of record $\mathbf{x}^*$, 
the remaining confidence score is uniformly spread over the rest classes. 
For example, consider a $10$-class classification problem, 
where the computed confidence score is $h(\mathbf{x}^*)=0.82$ and the label output represents Class 1. 
Then, the recovered confidence vector would be $\mathbf{y}=\langle 0.82,0.02,0.02,0.02,0.02,0.02,0.02,0.02,0.02,0.02\rangle$. 


\subsection{Step 5: train an attack model $A$}
Given a class of the target model and an auxiliary set, our method can reconstruct only one confidence vector. 
To create more confidence vectors to train the attack model, multiple sub-sets are sampled from the auxiliary set $D_{aux}$, 
where each sub-set is used to create a confidence vector following Steps 1-4. 
For each confidence vector created, a sample is randomly selected from the corresponding sub-set as the label. 
Then, these vectors, associated with their labels, are used to train the attack model $A$, 
where the $L_2$ loss function is used to train model $A$. 
After training, the attacker feeds the recovered confidence vector of $\mathbf{x}^*$ into model $A$ 
and receives the reconstructed data record $\mathbf{\widetilde{x}}^*$.


\section{Discussion of the attack method}\label{sec:analysis}

\subsection{Discussion of Steps 1 and 2}
In Steps 1 and 2, the data records are perturbed with Gaussian noise to test the target model $T$'s error rate. 
Then, the error rate is used to approximate the distance between the data record $\mathbf{x}^*$ and the decision boundary of $T$. 
This approximation is based on the following rationale. 

As shown in \cite{Fawzi16,Fawzi18}, for linear models, the error rate in a Gaussian distribution 
exactly determines the distance of the data records to the decision boundary. 
This means that, for a given data record $\mathbf{x}^*$, 
if we collect records based on the Gaussian distribution $\mathcal{N}(\mathbf{x}^*,\sigma^2\mathbf{I})$, 
then most records will lie close to the surface of a sphere of radius $\sigma$ centered at $\mathbf{x}^*$. 
Also, as the decision boundary of a linear model is a plane, 
the distance to the nearest error equals the distance from the plane to the center of the sphere $\mathbf{x}^*$. 
According to the analysis in \cite{Gilmer19}, this distance can be computed as $d(\mathbf{x}^*,E)=-\sigma\Phi^{-1}(\mu)$, 
where $E$ is the error set and $\mu$ is the error rate. 

However, in our problem, as the record $\mathbf{x}^*$ is unknown to the attacker, 
the distance $d(\mathbf{x}^*,E)$ cannot be computed directly. 
Thus, to approximately compute this distance, we instead collect records in the same class as $\mathbf{x}^*$, 
and add Gaussian noise $\mathcal{N}(0,\sigma^2\mathbf{I})$ to each of those records. 
According to Theorem \ref{thm:gaussian}, by adding the Gaussian noise, 
the median distance from one of the noisy records can be computed using the error rate. 
As this median distance is independent of any individual data record, 
we use the median distance with the auxiliary set $D_{aux}$ to approximate the distance $d(\mathbf{x}^*,E)$. 

It should be noted that the median distance $d^*_q(E)$ described in Theorem \ref{thm:gaussian} 
satisfies $d^*_q(E)\leq-\sigma\Phi^{-1}(\mu)$ when the error rate $\mu<\frac{1}{2}$ 
and the equality holds when $E$ is a half space. 
Therefore, to use $d^*_q(E)$ to approximate $d(\mathbf{x}^*,E)$, 
we must ensure that the error rate $\mu$ of the target model $T$ 
on the noisy dataset $\hat{D}_{aux}$ is less than $\frac{1}{2}$ 
by carefully controlling the amount of noise. This is done by tuning the value of $\sigma$. 
Moreover, to ensure an equality, i.e., $d^*_q(E)=-\sigma\Phi^{-1}(\mu)$, 
the error set $E$ must be a half space which means that the model must be linear. 
However, it is difficult to guarantee a linear target model $T$. 
Thus, using $d^*_q(E)$ to approximate $d(\mathbf{x}^*,E)$ will incur an error. 
To reduce this error, a linear shadow model $S$ is trained in Step 3 
and used to compute the distance $d(\mathbf{x}^*,E)$ by combining Eqs. \ref{eq:distance1} and \ref{eq:distance2}. 

\subsection{Discussion of Step 3}
A popular way to measure the performance of a model $f$ at predicting the label $y$ of a data record $\mathbf{x}$ 
is to use the log likelihood $log[p_f(y|\mathbf{x})]$ \cite{Zhang20}. 
This log likelihood, to a large extent, represents the classification error rate of the model $f$ on the label $y$. 
Therefore, to ensure the shadow model $S$ in Step 3 perform similarly to the target model $T$, 
the classification error rate of $S$ on the label of target record $\mathbf{x}^*$ must be close to 
the classification error rate of $T$ on the same label. 
Once the performance of the two models has been aligned, 
the distance from target record $\mathbf{x}^*$ to the decision boundary of model $T$ 
can be approximated by the distance from $\mathbf{x}^*$ to the decision boundary of $S$. 
Based on this distance, the confidence score $h(\mathbf{x}^*)$ can be computed using Eq. \ref{eq:confidence}.  
The deduction is given below.

By equating Eqs. \ref{eq:distance1} and \ref{eq:distance2}, we have 
\begin{equation}\nonumber
\begin{aligned}
    \frac{\alpha^{-1}(h(\mathbf{x}^*))}{||\mathbf{w}||_2}&=-\sigma\Phi^{-1}(\mu)\\
    \alpha^{-1}(h(\mathbf{x}^*))&=-\sigma\Phi^{-1}(\mu)||\mathbf{w}||_2.
\end{aligned}
\end{equation}
Given $\alpha(t)=\frac{1}{1+e^{-t}}$, we have 
\begin{equation}\nonumber
\begin{aligned}
\alpha^{-1}(t)&=ln\frac{t}{1-t}\\
\alpha^{-1}(h(\mathbf{x}^*))&=ln\frac{h(\mathbf{x}^*)}{1-h(\mathbf{x}^*)}=-\sigma\Phi^{-1}(\mu)||\mathbf{w}||_2\\
\Rightarrow h(\mathbf{x}^*)&=\frac{1}{1+exp[\sigma\Phi^{-1}(\mu)||\mathbf{w}||_2]}.
\end{aligned}
\end{equation}

\subsection{Discussion of Step 4}
Step 4 is a simple method of recovering the confidence scores for the remaining classes. 
As we will show in the experiments, even though this method is rudimentary, it is often quite effective. 
This is because the attack model, that is subsequently trained using the recovered confidence vectors, is not for classification. \
Instead, its goal is to reconstruct the general features of the given classes. 
Hence, the model is not very sensitive to a slight change in input. 
For example, given two confidence vectors $\mathbf{y}=(y_1,...,y_m)$ and $\mathbf{y}'=(y'_1,...,y'_m)$, 
if $argMax_{i\in[1,m]}y_i=argMax_{i\in[1,m]}y'_i$, then the two vectors $\mathbf{y}$ and $\mathbf{y}'$ indicate the same class. 
Therefore, as long as the attack model has a good generalization performance, 
its outputs should be very similar when taking $\mathbf{y}$ and $\mathbf{y}'$ as inputs.

Interestingly, we have also tested complex methods of recovering the remaining confidence scores: 
for example, taking the Euclidean distances between samples of different classes, 
where, intuitively, a smaller distance implies a larger confidence score. 
More concretely, assume that target model $T$ is a $10$-class classifier and the samples in $D_{aux}$ belong to the first class. 
Following Steps 1 to 3, we have $h(\textbf{x}^*)$. 
We then collect one sample from each of the remaining nine classes 
and use a sample from $D_{aux}$ to compute the Euclidean distance between this and the nine class samples. 
The resulting distances are then used to recover the remaining confidence scores. 
What we found was that overall performance with this more complicated method did not remarkably improve. 
We reasoned this might be because these samples are selected randomly. 
Thus, the Euclidean distance between two samples may not accurately reveal the relationship between two classes, 
which raises the question of a better method of estimating those relationships. 
Here, the class relationships in video categorization can be estimated using the weights of a trained model \cite{Jiang18}. 
However, in our problem, the weights are not available to the attacker. 
At this point, we decided to leave deep investigation of this issue to future work.

\subsection{Discussion of Step 5}
\begin{figure}[ht]
\centering
	\includegraphics[scale=0.45]{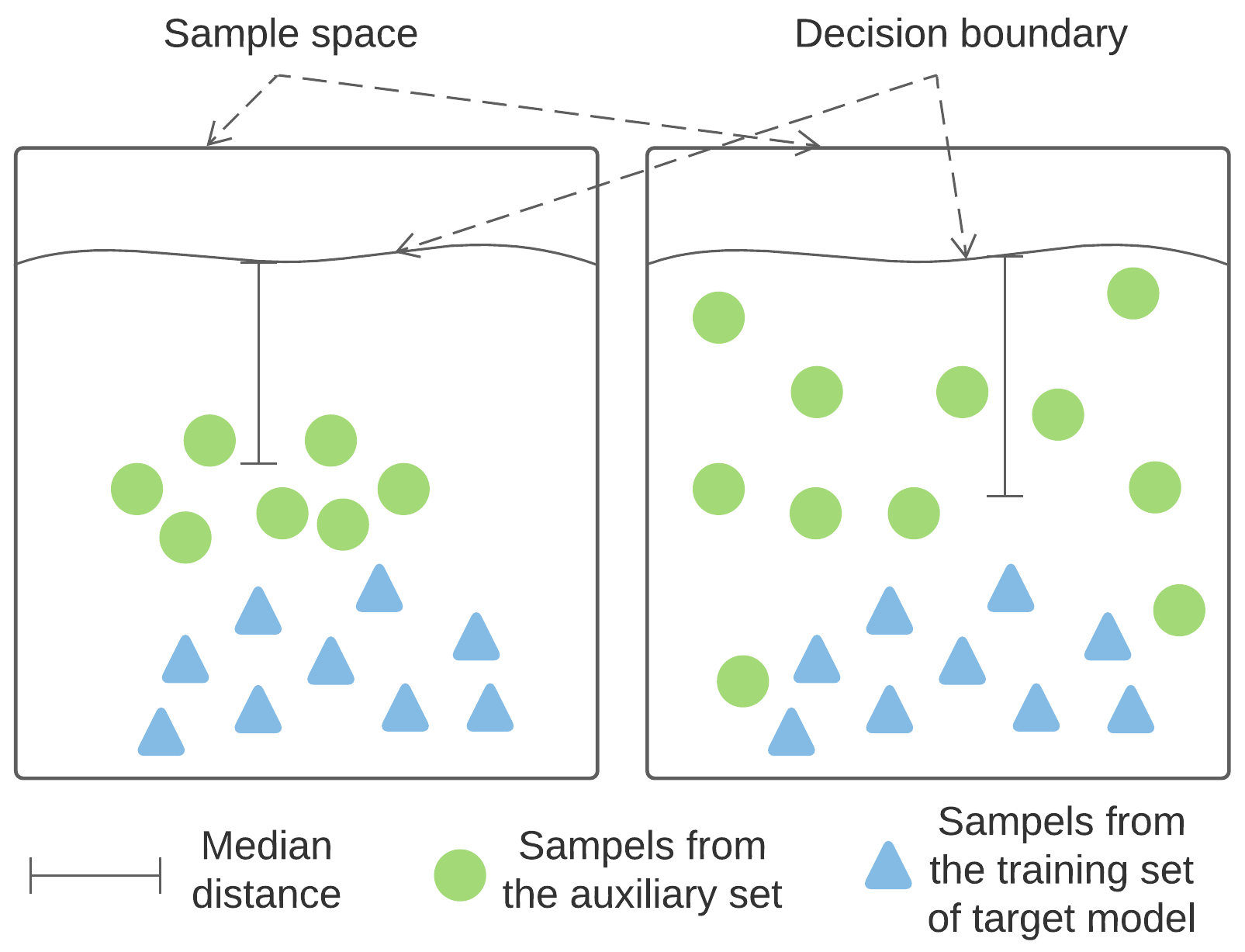}
	\caption{Median distance between the samples from an auxiliary set and the decision boundary of the target model. 
	The left figure shows a situation in which the auxiliary samples are concentrated, 
	while the right figure shows when they are separated. Both situations may arise.}
	\label{fig:Distance}
\end{figure}
Train the attack model is done with a set of vectors created by repeatedly sampling sub-sets from the auxiliary set and following Steps 1 to 4. 
Creating these multiple vectors is based on the following two considerations. 
First, intrinsically, using an auxiliary set to create a vector is based on 
the median distance between the samples in the auxiliary set and the decision boundary, as shown in the left panel of Figure \ref{fig:Distance}.
However, as the samples in the auxiliary set are collected arbitrarily, 
they can be distributed separately in the sample space, as shown in the right panel of Figure \ref{fig:Distance}. 
Hence, creating only one vector may not accurately estimate the distance 
between the given input $\mathbf{x}^*$ to the decision boundary of the target model. 
Moreover, as discussed in \cite{Li21}, samples not in the training set of the target model 
usually have a smaller distance to the decision boundary than the samples in the training set. 
Since the attacker does not know whether the auxiliary samples are in the training set of the target model, 
the best strategy is to use multiple sub-sets of the auxiliary set to create a set of confidence vectors. 

The second consideration is that, data augmentation is one of the popular methods 
of improving the attack model's generalization performance \cite{Neyshabur17}. 
This means the number of training samples, i.e., confidence vectors, for the attack model needs to be increased. 
In our attack, however, there is a challenge in training the attack model. 
Namely, the created confidence vectors do not have labels. 
To solve this challenge, as each confidence vector is generated based on a sub-set of the auxiliary set, 
a label is randomly selected from one of the samples in the corresponding sub-set. 
This is known as training on random labels \cite{Maennel20}. 
Typically, training on random labels increases the complexity of the trained model, which will then increase the generalization error \cite{Zhang17}. 
However, in our random label training scheme, there is a key difference to the typical approach. 
That is our random labels, i.e., the images coming from the auxiliary set, belong to one class, 
while in a typical approach, the labels are independent of each other 
and thus, randomizing the labels destroys any relationship between the data records and the labels. 
For this reason, the complexity of our attack model is less than the complexity of a model trained with a typical random-label training process. 
The empirical Rademacher complexity can explain this as follows. 
Given a hypothesis class $H$ on a data set $\{(\mathbf{u}_1,v_1);...;(\mathbf{u}_n,v_n)\}$, 
the empirical Rademacher complexity is defined as 
\begin{equation}\label{eq:Rad}
    Rad_n(H)=E[\mathop{sup}\limits_{f\in H}\frac{1}{n}\sum^n_{i=1}\sigma_i L(f(\mathbf{u}_i),v_i)],
\end{equation}
where $\sigma_1,...,\sigma_n\in\{-1,1\}$ are independent and uniform random variables. 
This definition introduces a randomization test that measures the ability of $H$ to fit a random $\pm 1$ binary label assignment. 
In our attack, the labels of attack model $A$ come from the auxiliary dataset $D_{aux}$. 
Thus, these labels belong to the same class. 
Since the distance between two samples that belong to the same class is smaller than that between two samples coming from different classes, 
if our training problem has the same hypothesis class as a typical random-label training problem, 
the empirical Rademacher complexity of our attack model is less than that of a model trained using random labels. 
Note that, to the best of our knowledge, training a model with correlated random labels is an open research issue. 
As such, in-depth research on this issue is left to future work. 

Another complex method we tried was to train the attack model using one generated confidence vector to match multiple samples from the auxiliary set. 
However, performance with this approach was worse than with the random label scheme. 
This kind of training is known as multi-label learning \cite{Read11,Read21}. 
Typically, the goal of multi-label learning is to find the ground-truth labels for a given set of data records. 
However, in our attack, a generated confidence vector does not have a ground-truth label, 
because the aim of the attack is to reconstruct a representative sample for each class. 
Therefore, any semantically meaningful sample can be the label of a given generated confidence vector. 
We refer to the training problem in our attack as ``fuzzy-label learning'', which is also an open research issue.

\section{Experiments}\label{sec:experiments}
We evaluate our attack method by comparison with three existing methods on four datasets. 
Four factors are evaluated in the experiments: 
the number of prediction classes, the size of the auxiliary set, the distribution of the auxiliary set and the structure of the attack models. 
We additionally quantified performance with two further metrics: the data inversion error and the confidence vector recovery error. 

\subsection{Experimental setup}
\subsubsection{Datasets}

\noindent\textbf{FaceScrub} \cite{Ng14} - a dataset of URLs for $100,000$ images of 530 individuals. 
    We collected $91,712$ images for $526$ individuals.
    Each image was resized to $64\times 64$.
    
\noindent\textbf{CelebA} \cite{Liu15} - a dataset with $202,599$ images of $10,177$ celebrities, i.e., classes, from the Internet. 
    After removing $296$ celebrities that are also included in FaceScrub, we were left with $195,727$ images of $9,881$ celebrities. 
    Again, each image was resized to $64\times 64$. 

\noindent\textbf{MNIST} \cite{LeCun98} - a dataset of $70,000$ images of handwritten numerals spanning $10$ classes: $0-9$. 
Each image was resized to $32\times 32$.
    
    
\noindent\textbf{CIFAR10} \cite{Krizhevsky14} - a dataset of $60,000$ images spanning $10$ classes, 
    including airplane, automobile, bird, cat, deer, dog, horse, ship and truck,  also resized to $32\times 32$.
    
Table \ref{tab:data} lists the data allocations for the experiments. 
To limit the attacker's knowledge, a very small part of each dataset was used to constitute the auxiliary set $D_{aux}$. 
$D_{aux}$ was further divided into two sets, $D^{train}_{aux}$ and $D^{test}_{aux}$, at a ratio of $9:1$ 
for training and testing the shadow model. 
Note that the ratio of samples, used to build the auxiliary sets for FaceScrub and CelebA, seems high at: $10\%$ and $25\%$, respectively, 
but when looking at the actual numbers of samples for each class, $10\%$ of FaceScrub equates to only about $20$ samples 
and $25\%$ of CelebA equates to only $5$ samples.

\begin{table}[!ht]\scriptsize
	\centering
	\caption{Data allocation}
\footnotesize
\begin{tabular}{|c|c|c|} \hline
\multicolumn{2}{|c|}{Target model $T$} & Shadow model $S$\\ \hline
Dataset & Data allocation & Auxiliary set \\ \hline
FaceScrub & $80\%$ train, $20\%$ test & $10\%$ of FaceScrub \\\hline
CelebA & $90\%$ train, $10\%$ test & $25\%$ of CelebA \\\hline
MNIST & $85\%$ train, $15\%$ test & $1\%$ of MNIST \\\hline
CIFAR10 & $80\%$ train, $20\%$ test & $1\%$ of CIFAR10  \\\hline
\end{tabular}
	\label{tab:data}
\end{table}

Moreover, as mentioned in the motivating example, we selected only irregular face images for the auxiliary samples for these two facial image datasets, 
e.g., left and right profiles. 
To select the images, we used the OpenCV library \cite{Bradski00} for Python3 and FAN network \cite{Bulat17} 
to detect the facial orientation of each image. 
Similarly, for the other two datasets, we hand-picked auxiliary samples with perceptibly fewer class features than the other samples. 

\subsubsection{Target and attack models}
We used the model architecture proposed in \cite{Yang19}. 
The FaceScrub, CelebA and CIFAR10 classifiers included 4 CNN blocks, two fully-connected layers and a softmax function. 
Each CNN block consisted of a convolutional layer followed by a batch normalization layer, 
a max-pooling layer and a ReLU activation layer. 
The two fully-connected layers were added after the CNN blocks. 
The softmax function was added to the last layer to convert neural signals into a valid confidence vector $\mathbf{y}$, 
where each $y_i\in [0,1]$ and $\sum^{k}_{i=1} y_i=1$, and $k$ is the number of classes. 
The attack models on the three datasets included 5 transposed CNN blocks. 
Each of the first 4 blocks had a transposed convolutional layer followed by a batch normalization layer and a $Tanh$ activation function. 
The last block had a transposed convolutional layer succeeded by a Sigmoid activation function 
that converted neural signals into real values in $[0,1]$. 
Similar to the FaceScrub configuration, the MNIST classifier had 3 CNN blocks and their attack models had 4 transposed CNN blocks.

 

\subsubsection{Comparison attack methods}
We compared our method, denoted as \emph{label-only}, with three other methods. 

\noindent\textbf{Vector-based} \cite{Yang19}: An attacker uses the output confidence score vectors of a target model 
to train their attack model for an inversion attack. 
This method has been experimentally proven to be the state-of-the-art in model inversion.

\noindent\textbf{Score-based} \cite{Yang19}: An attacker retains the highest confidence score in a score vector, 
while filling the rest of the scores with zeros. 
This attack is similar to a \emph{vector-based} attack but only one score is used. 

\noindent\textbf{One-hot}: This method is a simplified version of the \emph{label-only} method. 
Once the attacker receives the predicted class label, they simply set the confidence score of the predicted class to $1$ 
while setting the remaining scores to zeros. 

Although new methods have been proposed recently \cite{Salem20,Carlini21}, they are developed in different settings than ours. 
More explanation will be given in Section \ref{sec:related work}.

\subsubsection{Evaluation factors}


\noindent\textbf{The number of predicted classes}. 
Also referred to as prediction size, this is the number of classes that the target model can classify. 
This factor simulates real-world situations, where 
a target model may only be trained to classify some of these classes.

\noindent\textbf{The size of the auxiliary set}. 
This factor simulates real-world situations where 
different attackers may use different sized auxiliary sets to train their attack models.

\noindent\textbf{The distribution of the auxiliary set}. 
With this factor, we trained the target model using Facescrub but trained the attack model using CelebA. 
This simulates real-world situations where the attackers cannot collect samples 
of the same distribution as the target model's training set.

\noindent\textbf{The structure of the attack models}. Here, we removed a transposed CNN block from the attack models. 
This factor was used to evaluate the adaptability of the attack methods.

\subsubsection{Evaluation metrics}

\noindent\textbf{Data inversion error} shows the accuracy of the attack model at reconstructing the input data records. 
It is measured by computing the mean squared error (MSE) between an input data record and the reconstructed data record. 
Note that any input data records to the target model are actually unseen to the attacker. 
Here, we use them only for evaluating the experimental results. 

\noindent\textbf{Confidence vector recovery error} shows the accuracy of the attack model in recovering confidence vectors. 
It is measured by computing the Euclidian distance between a real confidence vector output by the target model and the recovered confidence vector. 
Real confidence vectors, actually, are unknown to the attacker. 
They are used only for evaluating the experimental results. 


\subsection{Experimental results}
\subsubsection{General inversion quality}
\begin{figure}[ht]
\centering
	\includegraphics[scale=0.26]{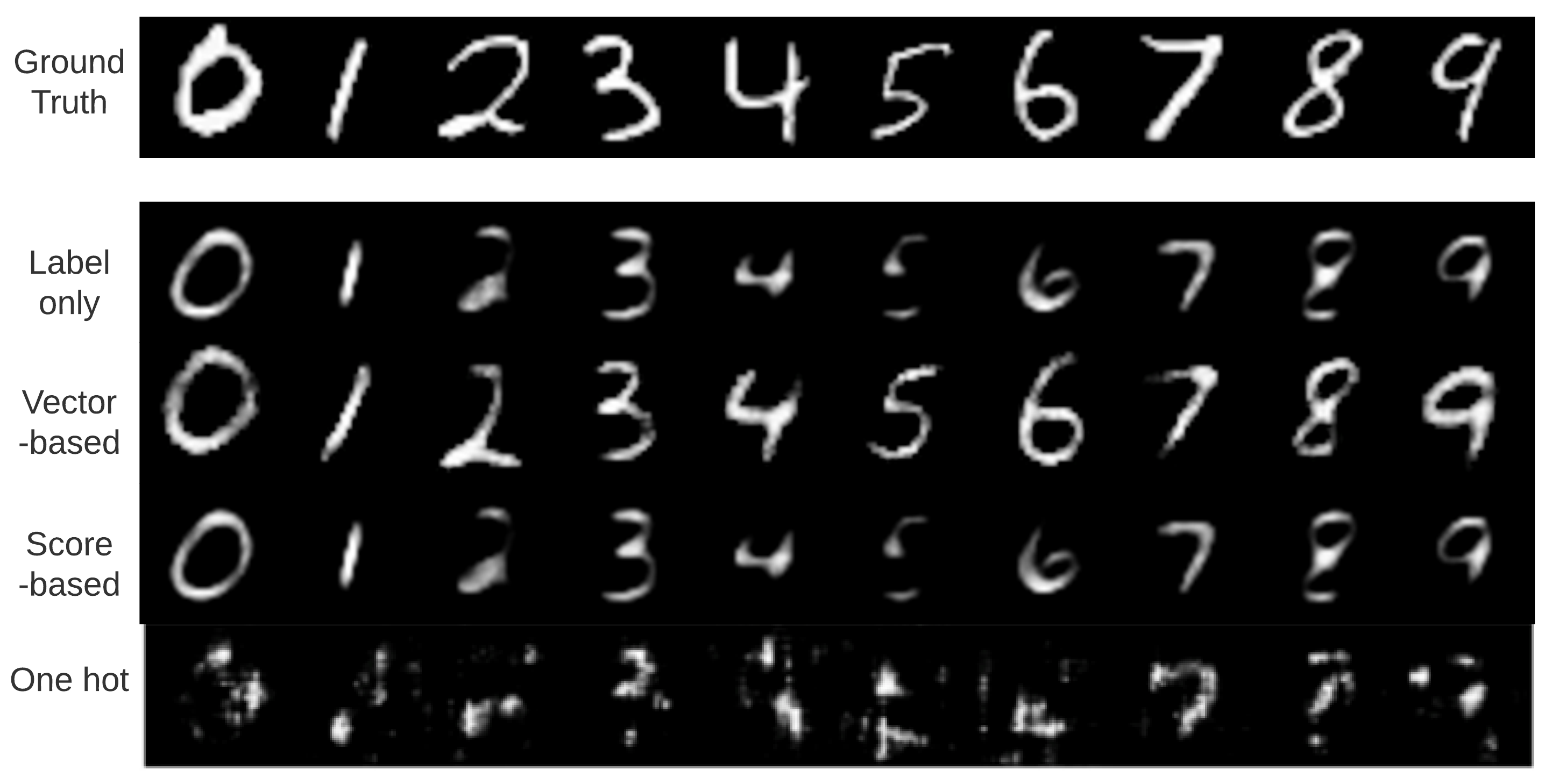}
	\caption{Performance of the four methods on MNIST} 
	\label{fig:MNIST}
\end{figure}


\begin{figure}[ht]
\centering
	\includegraphics[scale=0.30]{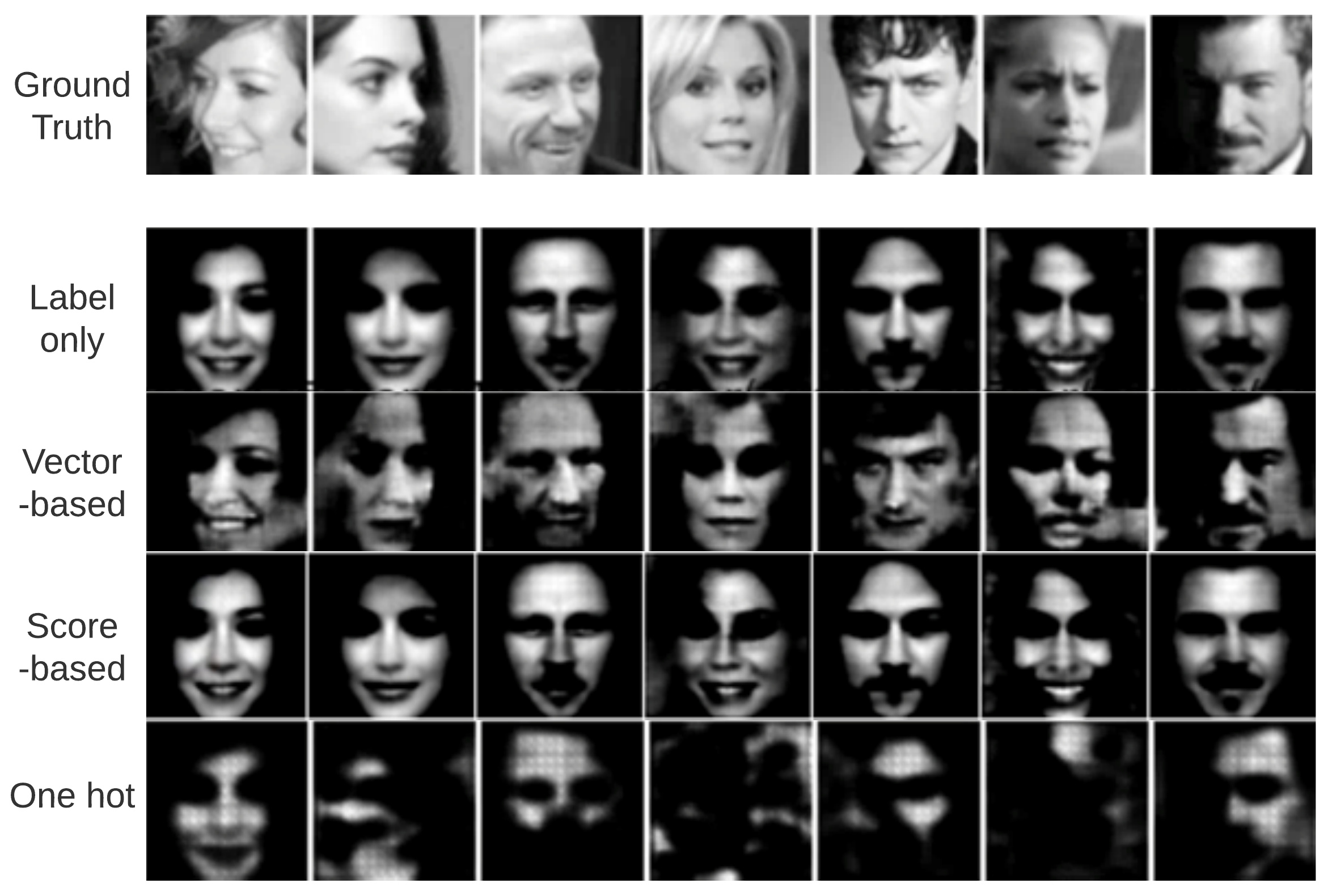}
	\caption{Performance of the four methods on FaceScrub} 
	\label{fig:FaceScrub}
\end{figure}

\begin{figure}[ht]
\centering
	\includegraphics[scale=0.37]{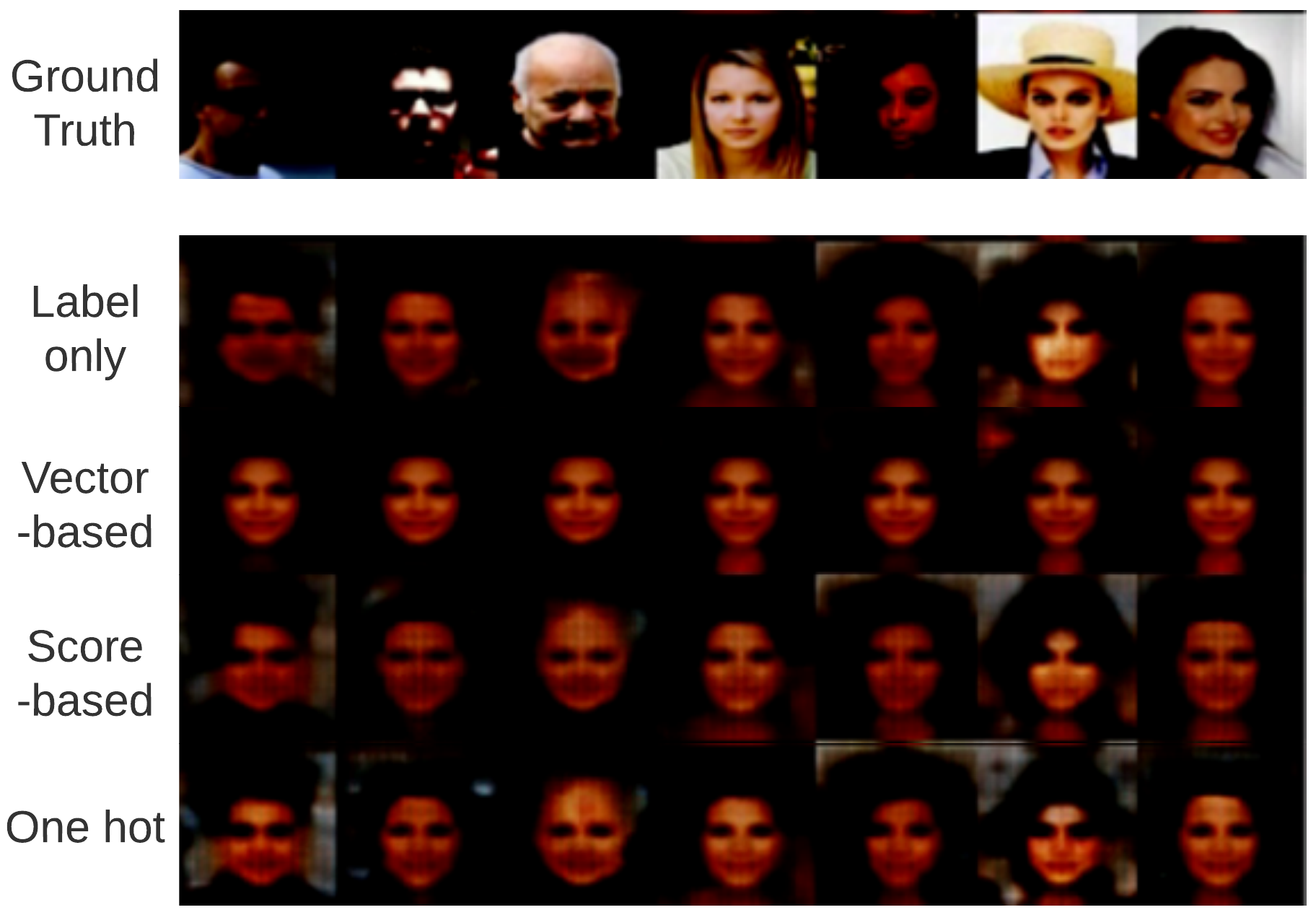}
	\caption{Performance of the four methods on CelebA} 
	\label{fig:CelebA}
\end{figure}

\begin{figure}[ht]
\centering
	\includegraphics[scale=0.37]{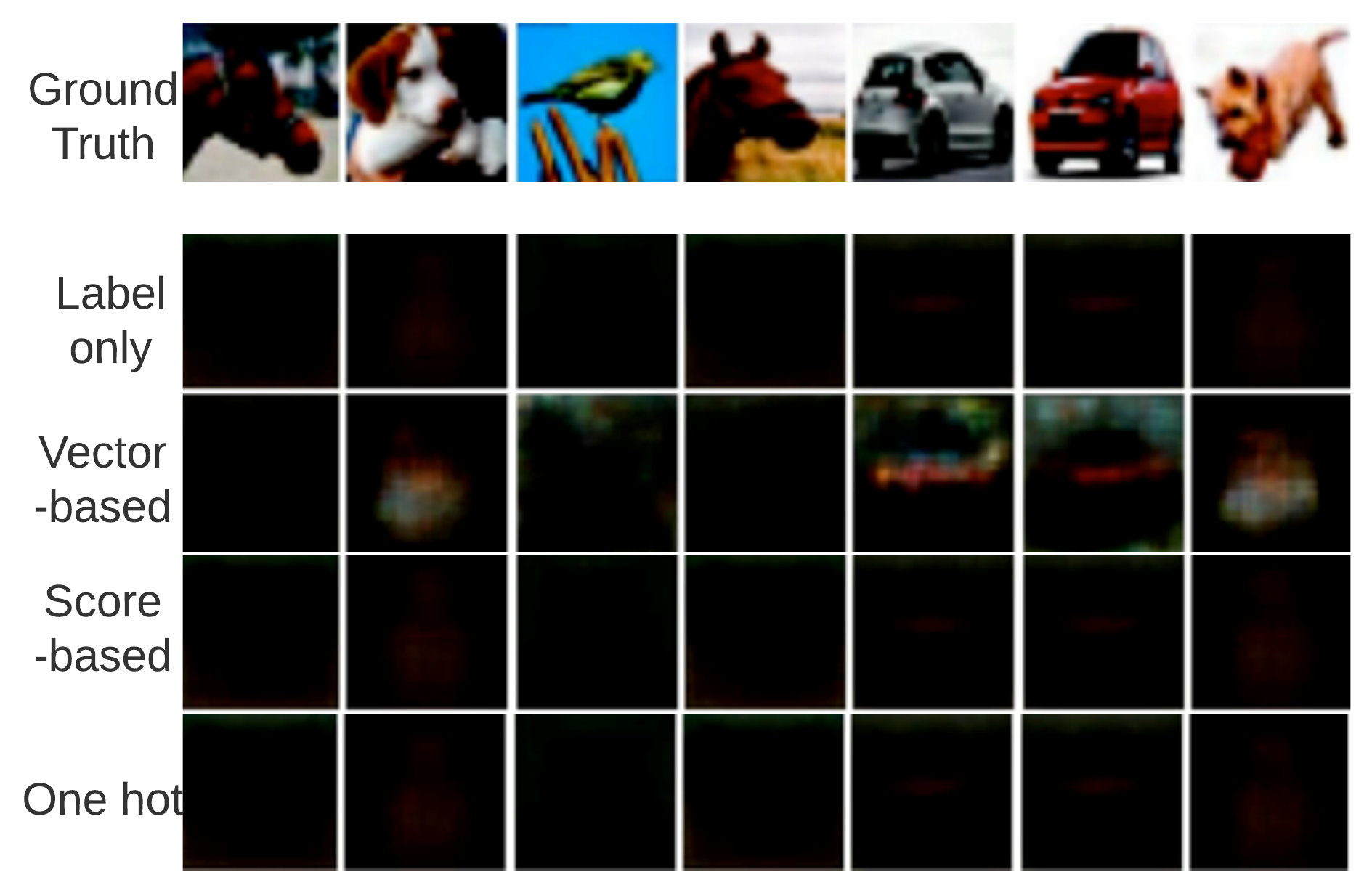}
	\caption{Performance of the four methods on CIFAR10} 
	\label{fig:CIFAR10}
\end{figure}

\begin{table}[!ht]\scriptsize
	\centering
	\caption{Data inversion errors of the four methods on the four datasets}
\begin{tabular}{|c|c|c|c|c|} \hline
& MNIST & FaceScrub & CelebA & CIFAR10 \\ \hline
Label-only & $0.873$ & $0.221$ & $0.362$ & $0.435$ \\ \hline
Vector-based & $0.841$ & $0.179$ & $0.301$ & $0.383$ \\ \hline
Score-based & $0.872$ & $0.219$ & $0.358$ & $0.432$ \\ \hline
One-hot & $0.968$ & $0.297$ & $0.416$ & $0.502$ \\ \hline
\end{tabular}
	\label{tab:Inversion}
\end{table}

\begin{table}[!ht]\scriptsize
	\centering
	\caption{Confidence vector recovery errors of the three methods on the four datasets}
\begin{tabular}{|c|c|c|c|c|} \hline
& MNIST & FaceScrub & CelebA & CIFAR10 \\ \hline
Label-only & $0.013$ & $0.051$ & $0.067$ & $0.073$ \\ \hline
Score-based & $0.010$ & $0.045$ & $0.063$ & $0.068$ \\ \hline
One-hot & $0.018$ & $0.065$ & $0.079$ & $0.095$ \\ \hline
\end{tabular}
	\label{tab:Recovery}
\end{table}

Figures \ref{fig:MNIST}, \ref{fig:FaceScrub}, \ref{fig:CelebA} and \ref{fig:CIFAR10} show 
the inversion quality of the four methods on MNIST, FaceScrub, CelebA and CIFAR10, respectively. 
Our \emph{label-only} method (the 2nd row in each figure) achieves almost the same result 
on all datasets as the \emph{score-based} method (the 4th row). 
Even though the only information available to the attacker with our method was the label of the input data record, 
the reconstructed images are still recognizable to humans on MNIST and FaceScrub. 
The \emph{vector-based} method (the 3rd row) achieves the best inversion quality 
due to the use of the rich information embodied in the confidence score vectors. 
The \emph{one-hot} method (the 5th row) delivers the worst inversion quality, 
as it simplifies each confidence score vector by setting the highest score to $1$ while masking the rest to $0$. 
This simplification largely destroys the information contained in each vector.  

This result is also supported by the quantitative data inversion errors as shown in Table \ref{tab:Inversion}, 
where the \emph{vector-based} method achieves the lowest inversion errors 
and the \emph{one-hot} method achieves the highest inversion errors. 
Moreover, in Table \ref{tab:Recovery}, we can see that there is a small gap between the real confidence vectors and the reconstructed confidence vectors. We reason that this is because the auxiliary samples we selected are of low-quality. Although the target model can correctly classify these auxiliary samples, its confidence in classifying them is low. Thus, in using these auxiliary samples to train the shadow models, there is inevitably a gap between the reconstructed confidence vectors and the real confidence vectors. 
Compared to the \emph{score-based} and \emph{one-hot} methods, 
we can see that the confidence vector recovery errors of our \emph{label-only} method 
are close to the \emph{score-based} method but smaller than the \emph{one-hot} method. 
This also explains the fact that the performance of our \emph{label-only} method is similar to the \emph{score-based} method 
and much better than the \emph{one-hot} method. 
Note that in Table \ref{tab:Inversion}, as all samples within one class have the same label, 
the inversion errors in the rows aligned to our method and the \emph{one-hot} method are obtained 
by averaging the mean squared errors between a reconstructed record and a randomly selected batch of the original records. 
Likewise, in Table \ref{tab:Recovery}, the vector recovery errors of the three methods are computed 
by averaging the recovery errors of a randomly selected set of test samples.

Specifically, in Figure \ref{fig:MNIST}, for the digits `2', `4' and `5', 
the inversion quality of the \emph{vector-based} method is better than the \emph{label-only} and \emph{score-based} methods. 
This is because the information contained in a label is only enough to reconstruct generic features of an image. 
In our \emph{label-only} method, only the highest score is recovered, while the remaining score is uniformly spread to the rest classes. 
Technically, the highest score contains the generic features of an image, while the remaining score embodies details of an image. 
Thus, the details of an image are hard to recover with our \emph{label-only} method. 
This theory also holds for the \emph{score-based} method, 
where only the highest score is used by the attacker and the rest of the scores are filled with zeros. 


\begin{figure}[ht]
\centering
	\includegraphics[scale=0.29]{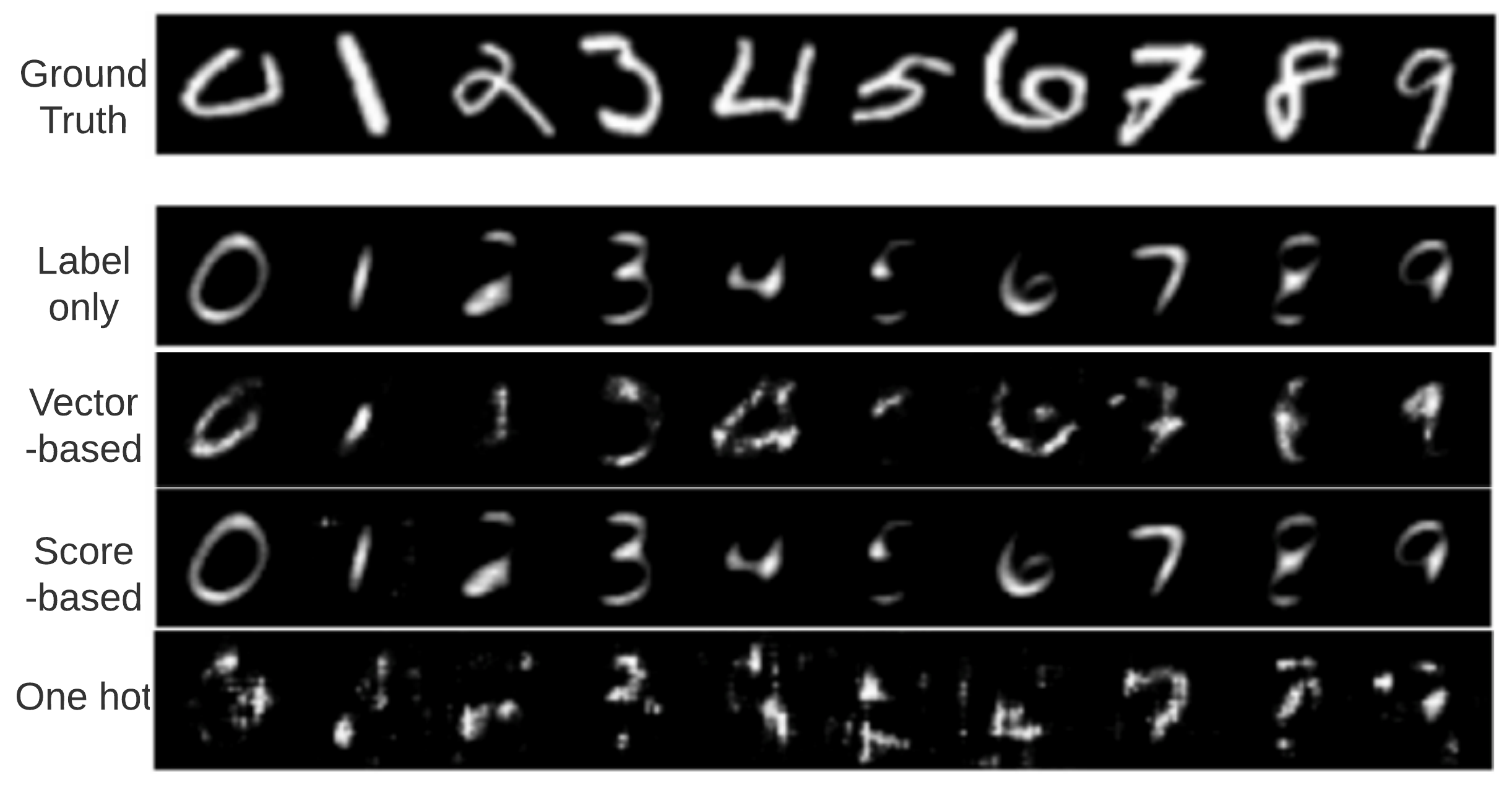}
	\caption{Performance of the three methods on another ground truth set from MNIST} 
	\label{fig:AnotherMNIST}
\end{figure}

However, we find that the inversion quality of the \emph{vector-based} method highly depends on the given data records. 
For example, in Figure \ref{fig:AnotherMNIST}, when we use another set of data records from MNIST as the ground truth, 
the inversion quality of the \emph{vector-based} method turns out to be worse than the \emph{label-only} and \emph{score-based} methods. 
This is because the \emph{vector-based} method aims to reconstruct every detail of a given data record. 
Hence, if the quality of the given record is not good, the quality of the reconstructed record will also not be good. 
By comparison, the \emph{label-only} and \emph{score-based} methods aim to recover the generic features of a given data record. 
Thereby, even if the quality of the given record is not good, 
its generic features can still be recovered. 

This finding can also be used to explain the results in Figure \ref{fig:FaceScrub}. 
When the input facial images contain only profiles, 
the images recovered by the \emph{vector-based} method have unexpected patterns (see the first and second columns). 
In contrast, the \emph{label-only} and \emph{score-based} methods are able to clearly reconstruct the general features of these images. 
These results are significant to situations where 
an attacker can only collect profile facial images of a person but they are curious as to what this person looks like.

Unfortunately, none of the methods work well on CelebA and CIFAR10, as shown in Figures \ref{fig:CelebA} and \ref{fig:CIFAR10}, respectively. 
For CelebA, we attribute the poor result to the small number of samples for each class at about $20$. 
This means the target model itself could not be trained well. 
With all four attack methods, good performance relies on the output of the target model 
and all four attack methods only manage to reconstruct ``average'' faces, 
where eyes, noses and mouths are recognizable but the gender or actual identity are indiscernible. 
This indicates that the attack models are trained incorrectly, probably as a result of a poorly trained target model. 
For CIFAR10, most images contain redundant background information, e.g., blue sky, white cloud, green grass, etc.. 
This kind of information is difficult to recover using only a vector.

\subsubsection{Effects of the prediction size}
\begin{figure}[ht]
\centering
	\includegraphics[scale=0.36]{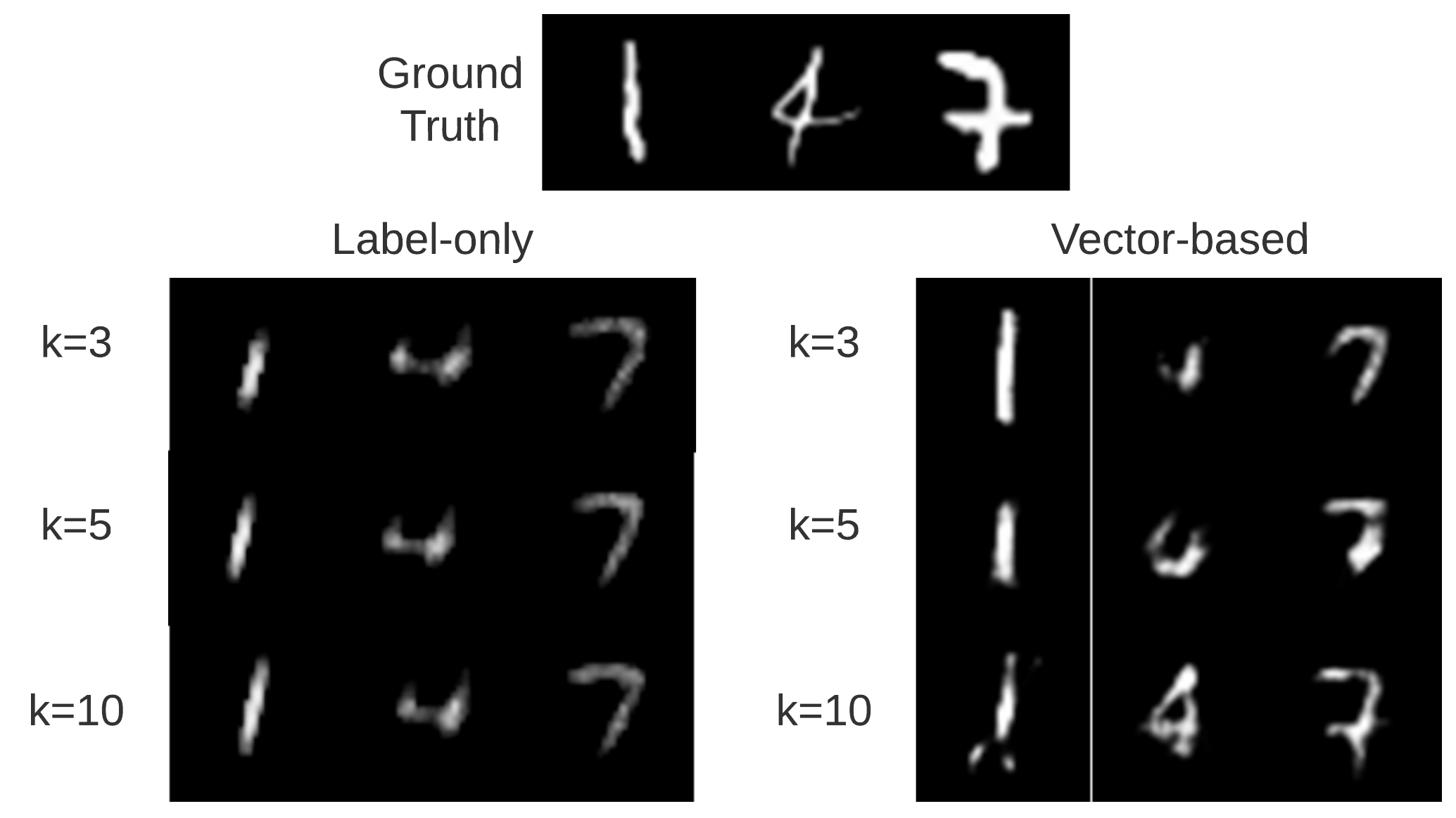}
	\caption{Performance of the \emph{label-only} and \emph{vector-based} methods on MNIST with different numbers of classes.
	Performance is evaluated using data records for $3$ classes, $5$ classes and $10$ classes.}
	\label{fig:MNIST-Class}
\end{figure}


\begin{figure}[ht]
\centering
	\includegraphics[scale=0.42]{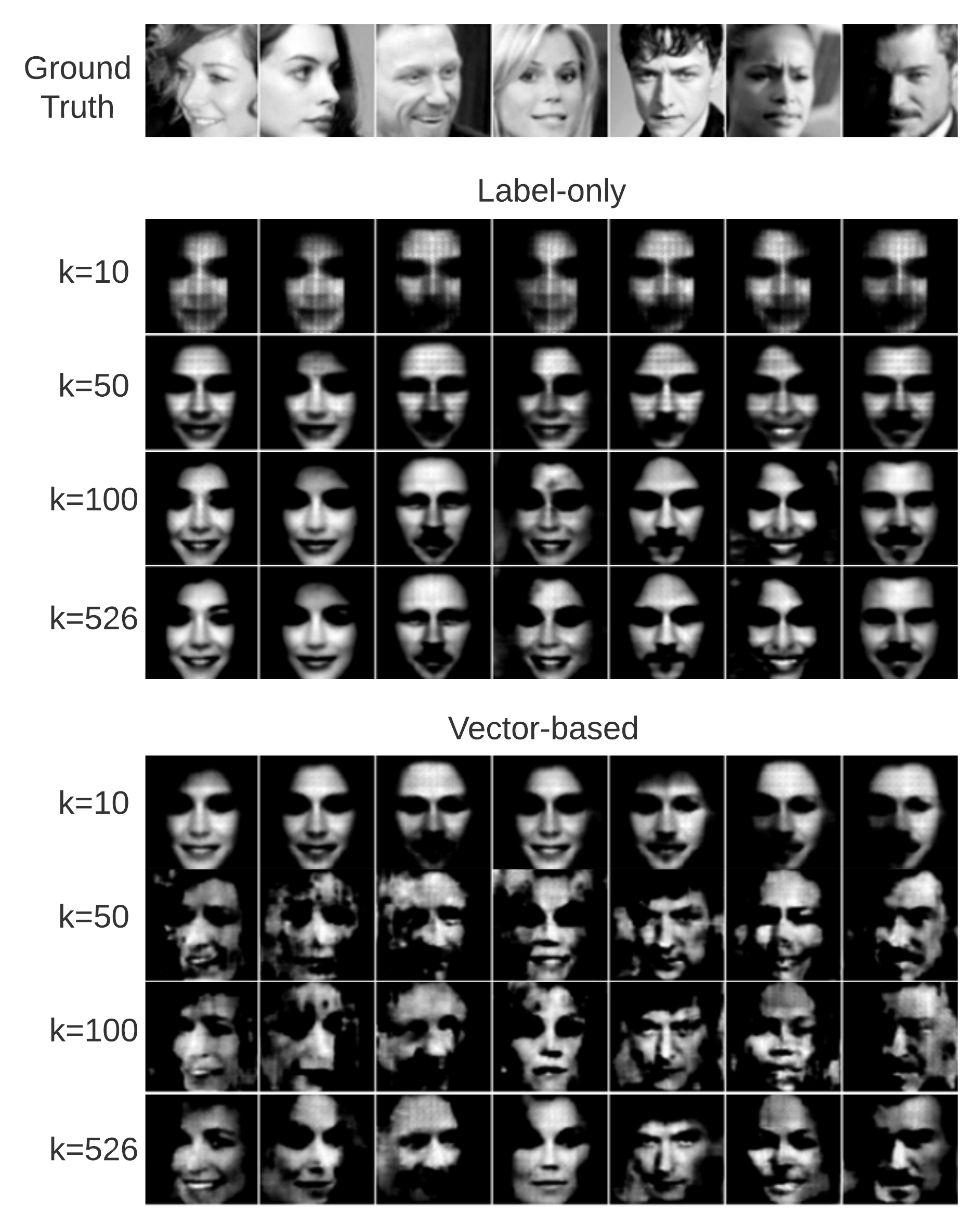}
	\caption{Performance of the \emph{label-only} and \emph{vector-based} methods on FaceScrub with different numbers of classes.
	Performance is evaluated using data records for $10$ classes, $50$ classes, $100$ classes and $526$ classes.}
	\label{fig:FaceScrub-Class}
\end{figure}

In the remaining experiments, the \emph{score-based} method performs similarly to the \emph{label-only} method, and the \emph{one-hot} method generally fails to reach good levels of recognizability. Hence, due to space limitations, we have moved the results for these two methods to the Appendix.
The tables listing data inversion errors and confidence vector recovery errors can also been found in the Appendix. 
Lastly, since no method performs well on CelebA or CIFAR10, the corresponding results are not presented. 

Figures \ref{fig:MNIST-Class} and \ref{fig:FaceScrub-Class} show the inversion quality of the \emph{label-only} 
and \emph{vector-based} methods on MNIST and FaceScrub, respectively, with different numbers of classes. 
The overall results demonstrate that the inversion quality of our method is independent of the number of classes used for training. 
This can be explained by the fact that our method mainly uses the highest confidence score to do the inversion. 
Therefore, the number of classes, i.e., the number of scores in a confidence vector, does not have much impact on the performance of our method. 
In contrast, the inversion quality of the \emph{vector-based} method highly depends on the number of classes used for training attack models, 
because the \emph{vector-based} method uses all the scores in the confidence vectors to reconstruct the data records. 
Hence, if some of the scores are missing, the performance of the \emph{vector-based} method is significantly affected. 
An interesting finding is that, in Figure \ref{fig:FaceScrub-Class}, when $k=10$, 
similar to our \emph{label-only} method, the \emph{vector-based} method reconstructs only the general features of the facial images. 
This result reveals the fact that if the amount of predicted information is reduced, 
even the most powerful attacker can do only a generic inversion. 


\subsubsection{Effects of the auxiliary set size}
\begin{figure}[ht]
\centering
	\includegraphics[scale=0.24]{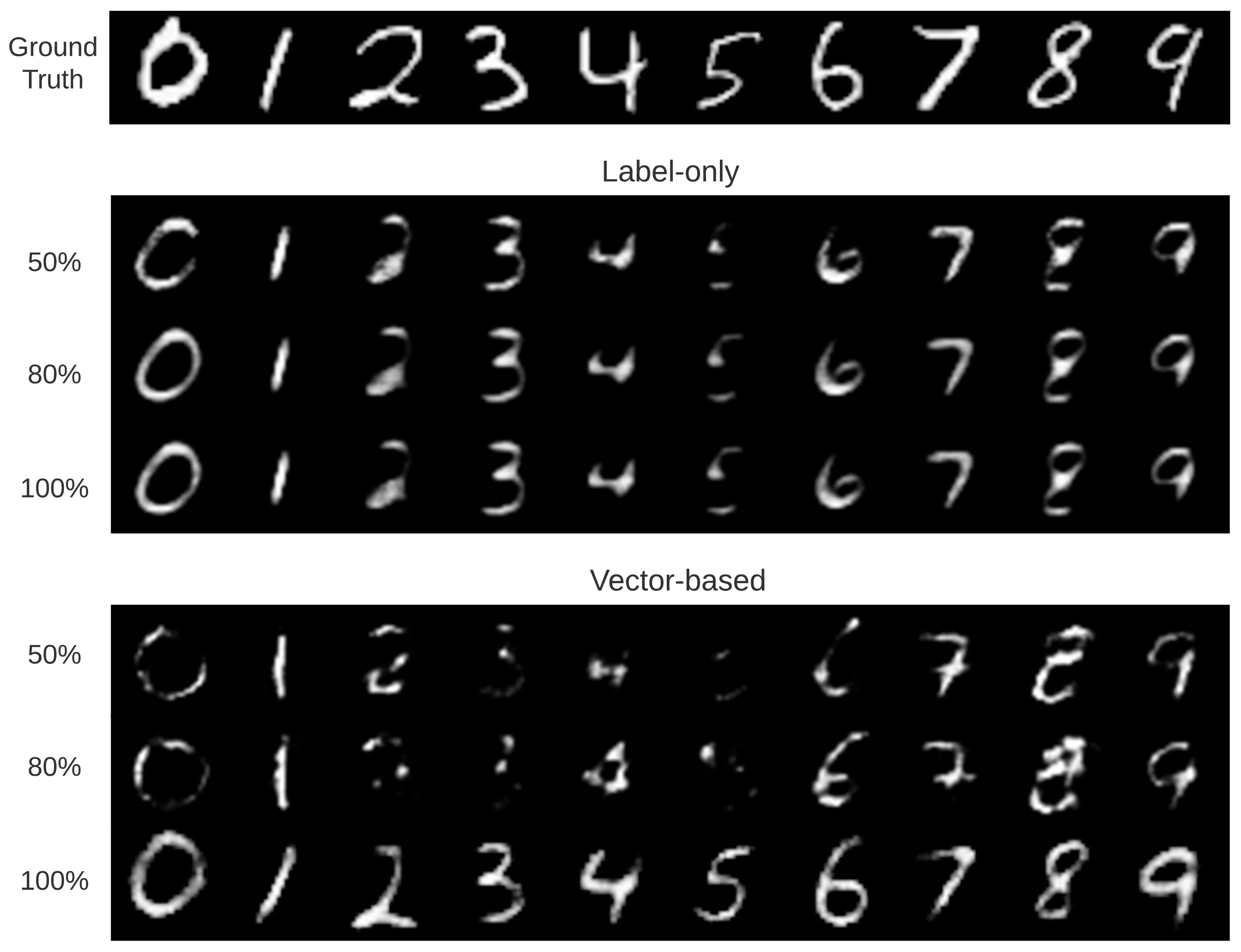}
	\caption{Performance of the \emph{label-only} and \emph{vector-based} methods on MNIST with different sized auxiliary set. 
	Performance is evaluated using $50\%$, $80\%$ and $100\%$ of the auxiliary set.}
	\label{fig:MNIST-Set}
\end{figure}


\begin{figure}[ht]
\centering
	\includegraphics[scale=0.27]{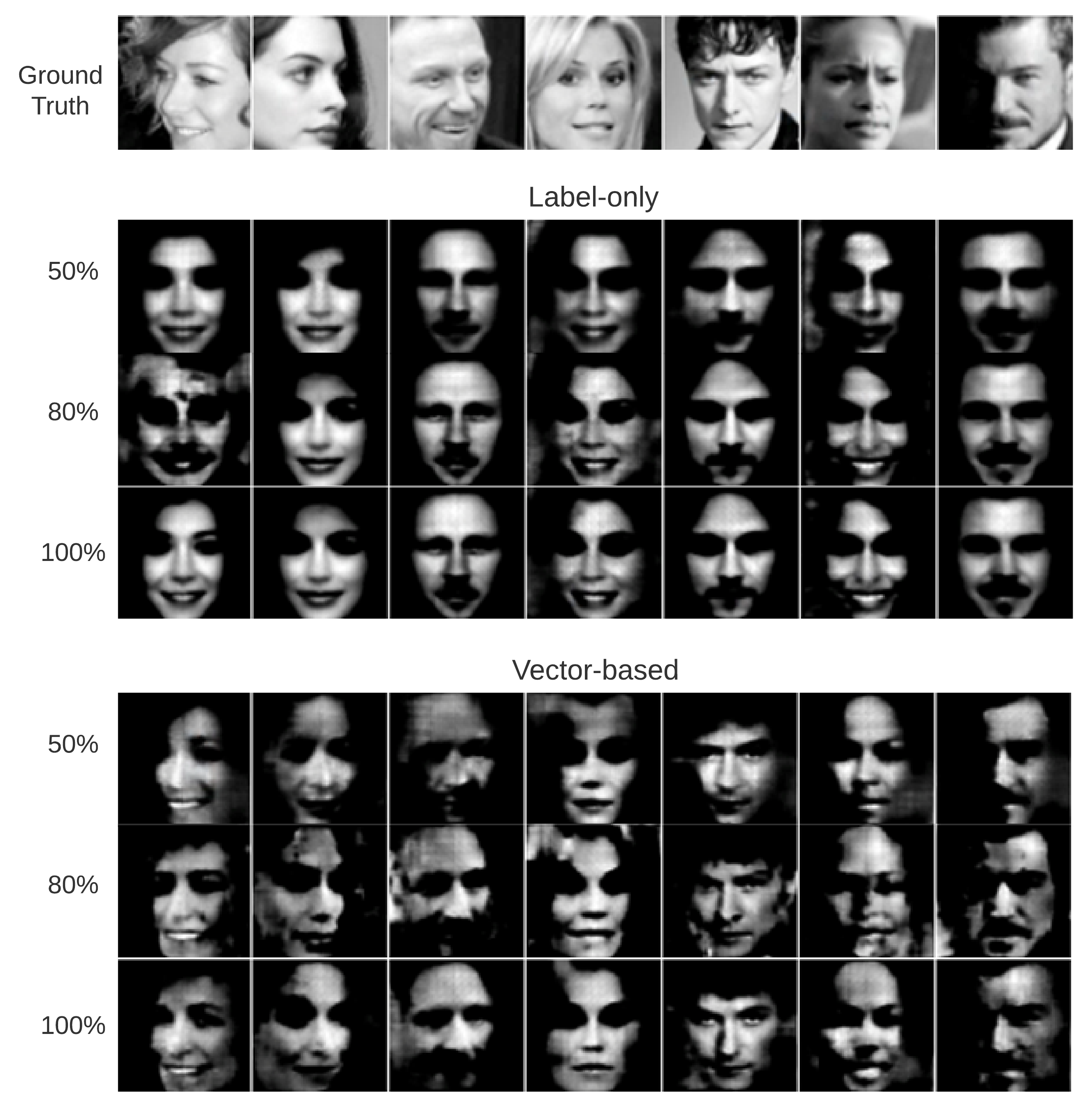}
	\caption{Performance of the \emph{label-only} and \emph{vector-based} methods on FaceScrub with a different sized auxiliary set. 
	Performance is evaluated using $50\%$, $80\%$ and $100\%$ of the auxiliary set.}
	\label{fig:FaceScrub-Set}
\end{figure}

Figures \ref{fig:MNIST-Set} and \ref{fig:FaceScrub-Set} show the inversion quality 
with different sized auxiliary sets for the \emph{label-only} and \emph{vector-based} methods on MNIST and FaceScrub, respectively. 
It can be seen that the size of the auxiliary set has very little impact on the inversion quality of our method, 
but it does have significant impact on the \emph{vector-based} method, especially with MNIST. 
This is because the two methods have different learning trends. 
Given a data record, our \emph{label-only} method learns the generic features of the set of data records 
that belong to the same class as the given record, 
while the \emph{vector-based} method learns to recover the specific features of the given record. 
Thus, a few samples are sufficient to train a good attack model in our \emph{label-only} method, 
but are not enough for the \emph{vector-based} method.


Specifically, a close look at Figure \ref{fig:MNIST-Set} reveals that 
the inversion quality of our \emph{label-only} method when using $50\%$ of the auxiliary set is a bit worse 
than using $80\%$ or $100\%$ of the set, 
where slightly more black-and-white noise can be seen on the digits. 
The results for Figure \ref{fig:FaceScrub-Set} are similar, 
where using $80\%$ and $100\%$ of the auxiliary set achieves a slightly better inversion quality than only using $50\%$. 

For the \emph{vector-based} method, in Figure \ref{fig:MNIST-Set}, the inversion quality of digits `3', `4' and `5' becomes worse 
when the size of the auxiliary set is reduced. 
Also, in Figure \ref{fig:FaceScrub-Set}, Columns 3 and 4, when $50\%$ of the auxiliary set is used, 
large black patterns appear in the recovered images. 
This means that the \emph{vector-based} method finds it difficult to reconstruct specific features of the given records with a small auxiliary set.


\subsubsection{Effects of the auxiliary set distribution}
We use FaceScrub to train the target model and CelebA as the auxiliary set to train attack models. 
In addition, we ensure that CelebA does not have a class intersection with FaceScrub.  
Thus, the attack models have never seen the classes of the target model during their construction. 
To train the attack models, we feed CelebA samples into the target model and use its output as training samples for attack models.

\begin{figure}[ht]
\centering
	\includegraphics[scale=0.4]{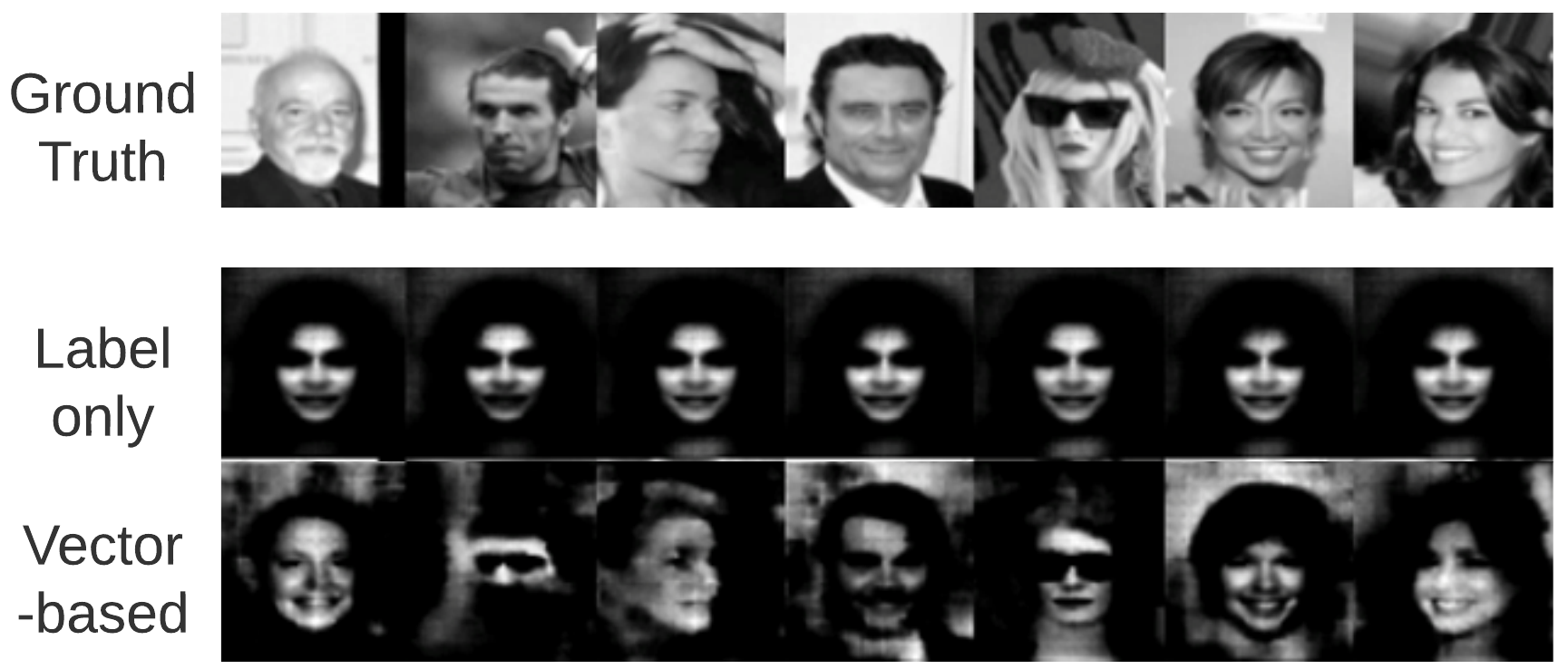}
	\caption{Performance of the \emph{label-only} and \emph{vector-based} methods when using FaceScrub to train the target model and CelebA to train the attack model.}
	\label{fig:FaceScrub-distribution}
\end{figure}

The results are shown in Figure \ref{fig:FaceScrub-distribution}. 
We can see that the \emph{vector-based} method is able to reasonably recover input images, 
while our \emph{label-only} method is only able to reconstruct ``average'' faces, 
where all the reconstructed faces look similar irrespective of their class. 
We think this may be because our method requires the error rates of the target model to generate confidence scores. 
The auxiliary samples from a generic distribution do not belong to any class of the target model. 
Hence, the target model cannot give any meaningful error rates on these auxiliary samples. 
Thus, the generated confidence scores may not approximate the real confidence scores, 
which yields results that do not contain the general features of the training samples. 
One potential way to improve the quality of the results would be to investigate the relationship 
between the classes of the auxiliary set and the classes of the training set. 
Then, when combined with the error rates, the relationships could be used to generate meaningful confidence scores. 
We leave this as our future work.

\subsubsection{Effects of attack model structure}
\begin{figure}[ht]
\centering
	\includegraphics[scale=0.35]{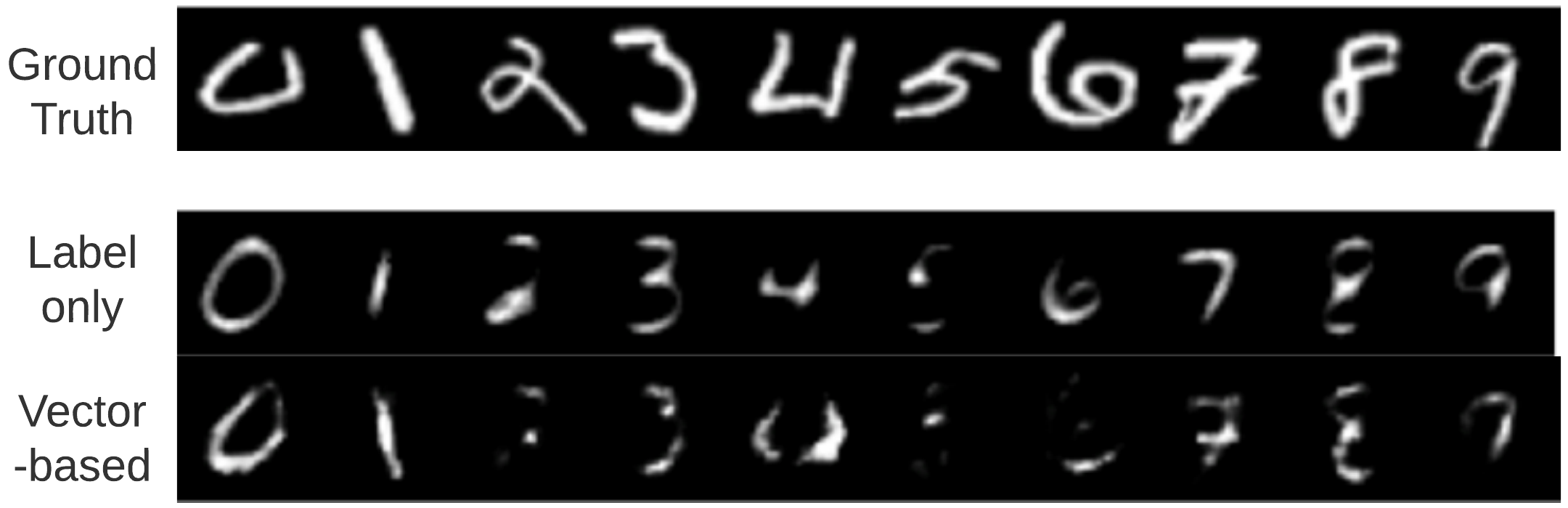}
	\caption{Performance of the \emph{label-only} and \emph{vector-based} methods on MNIST when removing a transposed CNN block from attack models.}
	\label{fig:MNIST-structure}
\end{figure}

\begin{figure}[ht]
\centering
	\includegraphics[scale=0.40]{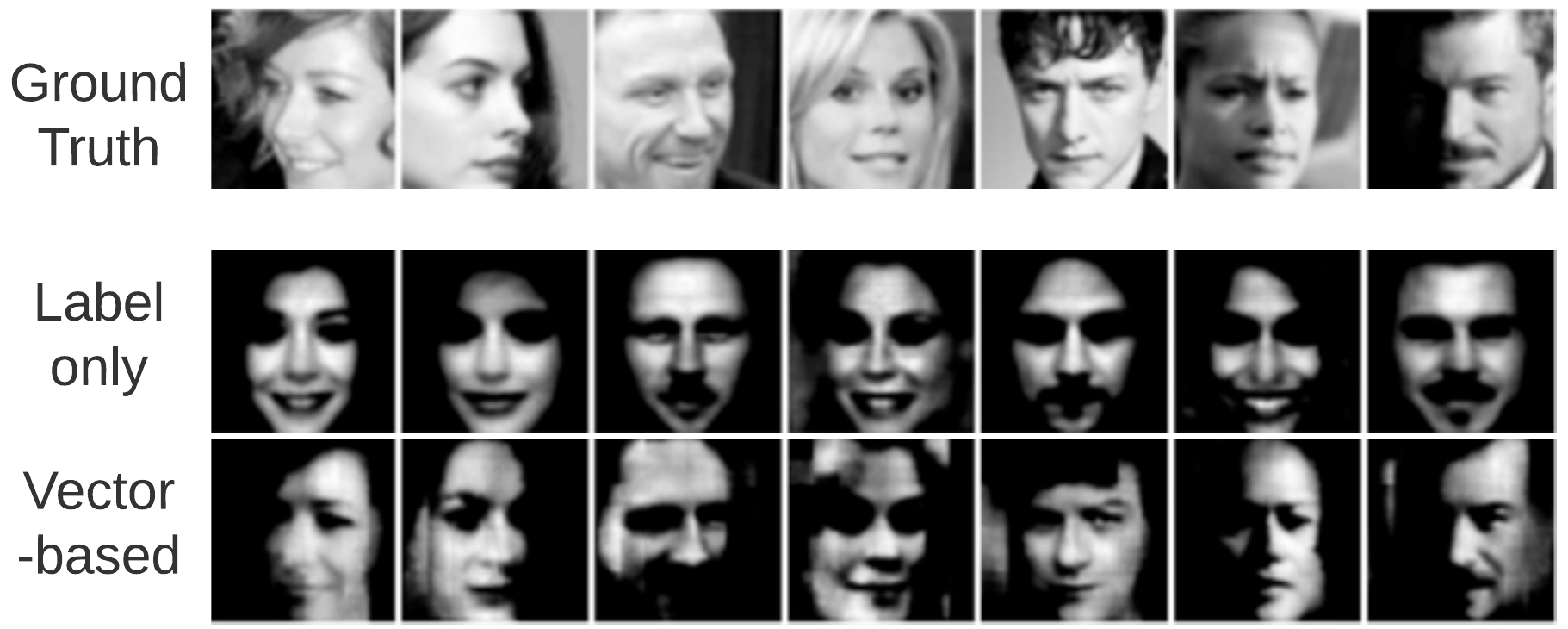}
	\caption{Performance of the \emph{label-only} and \emph{vector-based} methods on FaceScrub when removing a transposed CNN block from attack models.}
	\label{fig:FaceScrub-structure}
\end{figure}

We removed a transposed CNN block from the attack models. 
In Figures \ref{fig:MNIST-structure} and \ref{fig:FaceScrub-structure}, 
we can see that our \emph{label-only} method can still reconstruct the generic features of each class 
whereas the performance of the \emph{vector-based} method is affected to some extent. 
This result is because the \emph{vector-based} method aims to precisely recover each sample input to the target model. 
Thus, its performance highly depends on the hyper-parameters of the attack model, including its structure. 
By comparison, our method aims to reconstruct generic features of each class. 
Hence, its performance is not as sensitive to the attack model structure as the \emph{vector-based} method. 

\subsection{Summary of experiments}
According to the experimental results, our \emph{label-only} method can achieve the same performance as the \emph{score-based} method 
and outperforms the \emph{one-hot} method, but is not as good as the \emph{vector-based} method. 
This is because the only information used by our method is the label of an input data record. 
However, the performance of the \emph{vector-based} method highly depends on 
the given records, the prediction size, the size of the auxiliary sets and the structure of the attack models. 
In contrast, as our method only requires a small amount of information and that information is independent of the prediction size, 
the auxiliary set size and the attack model structure, our method has greater adaptability than the \emph{vector-based} method.

\section{A potential defense method}\label{sec:defense}
Since our attack method only uses the output labels of the target model, the popular defense methods \cite{Shokri17,Jia19,Yang20}, 
which modify the output confidence vectors, would do little to combat our attack method. 
A potential defense method might be to disturb the output labels. 
The rationale is that if the target model can intentionally modify the output labels, 
then the attacker cannot effectively use the target model's error rates to compute confidence vectors. 
Hence, the success of the attack might be compromised. 
However, disturbing the output labels will inevitably introduce classification errors into the target model. 
Given that each classifier has an intrinsic classification error when classifying a given dataset, 
we can limit the introduced classification error to less than the intrinsic error.
Based on this idea, we develop a defense method. 
Given an input data record, the target model, with a probability of $p$, randomly allocates a label to the input record. 
The probability $p$ is a hyper-parameter whose value should be less than the target model's intrinsic classification error rate $p^*$.

\begin{figure}[ht]
\centering
	\includegraphics[scale=0.27]{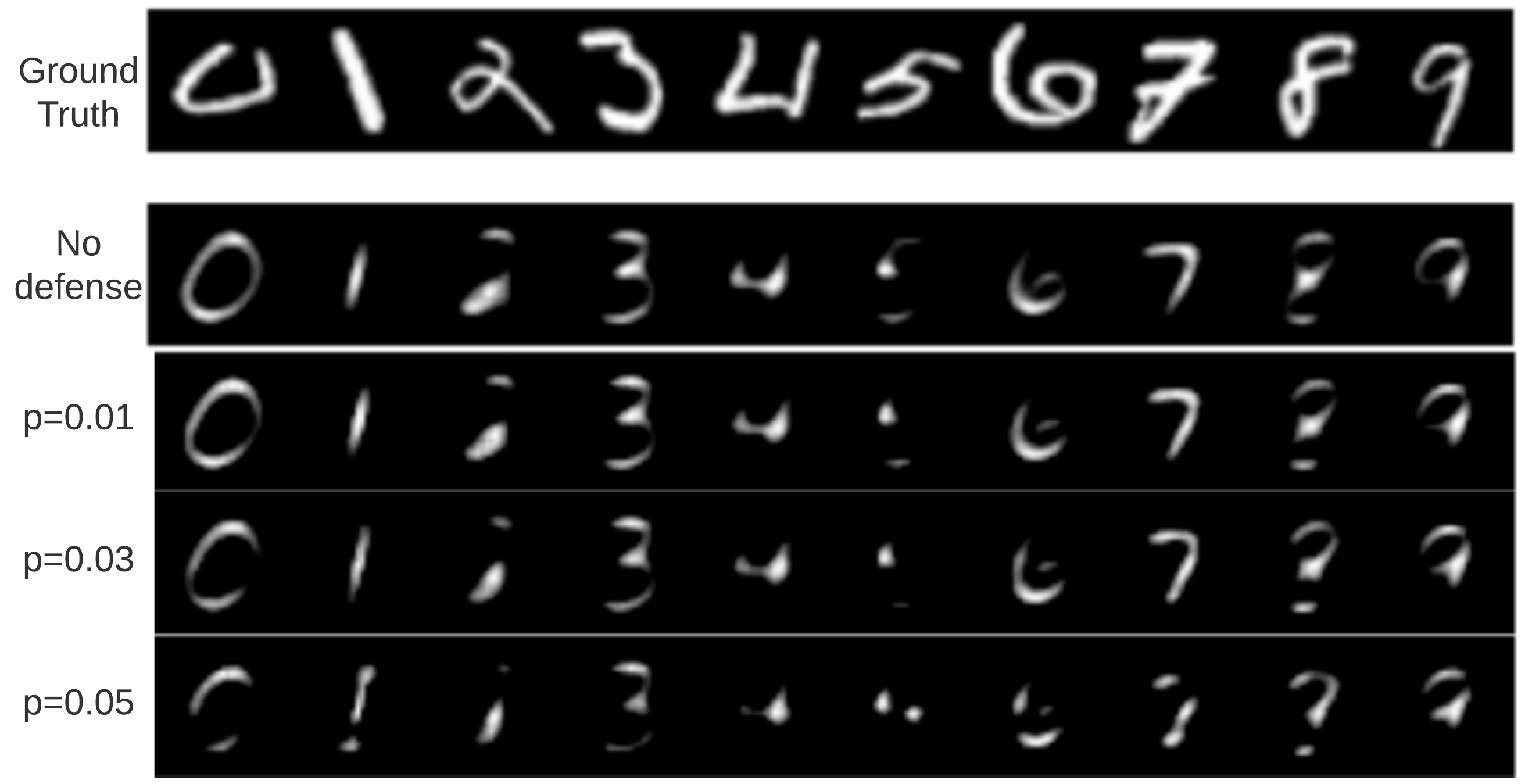}
	\caption{Defense against the \emph{label-only} method on MNIST by setting the error rate to $0.01$, $0.03$ and $0.05$.}
	\label{fig:MNIST-Defense}
\end{figure}

\begin{figure}[ht]
\centering
	\includegraphics[scale=0.27]{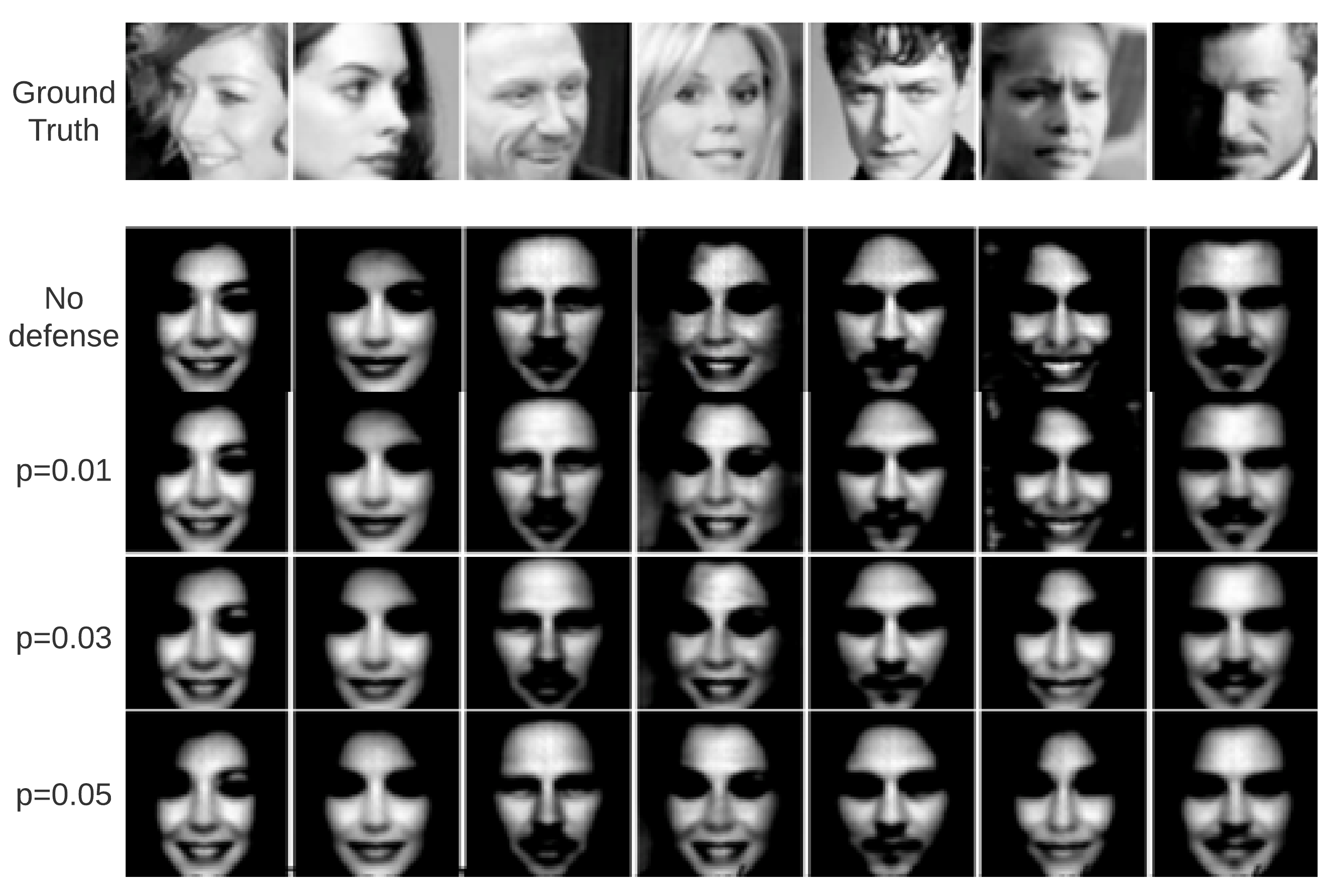}
	\caption{Defense against the \emph{label-only} method on FaceScrub by setting the error rate to $0.01$, $0.03$ and $0.05$.}
	\label{fig:FaceScrub-Defense}
\end{figure}

Applying this defense method in our experiments, 
we find that as the introduced error rate $p$ increases, the overall inversion quality decreases as shown in Figures \ref{fig:MNIST-Defense} and \ref{fig:FaceScrub-Defense}. 
For example, in Figure \ref{fig:MNIST-Defense}, when $p=0.05$, the digits `2', `5', `6' and `8' are hard to identify. 
Also, in Figure \ref{fig:FaceScrub-Defense}, when $p=0.05$, in the $7$th column, the man's beard disappears. 
The results show that the defense method is effective in some situations but not quite effective in others. 
We leave the thorough research of the defense method as our future work.

\section{Related work}\label{sec:related work}

\subsection{Black-box model inversion attacks}
Fredrikson et al. \cite{Fred14} initiated the research of model inversion attacks 
on personalized warfarin dosing in a black-box manner. 
The adversary is given access to a target model, the warfarin dosage of an individual, 
and some rudimentary information about the dataset. 
The adversary aims to predict one of the genotype attributes for that individual. 
Their method works by estimating the probability of a potential target attribute 
given the available information and the model. 

Later, Fredrikson et al. \cite{Fred15} generalized the work in \cite{Fred14} to both black-box and white-box scenarios 
by proposing a new class of model inversion attacks that 
exploit the confidence scores revealed along with the prediction. 
Their attack method is based on decision trees 
that access the model and side information from the data. 
In return, the attacker receives confidence measures for each of the classifications. 

Hidano et al. \cite{Hidano17} improved on Fredriskon et al's work \cite{Fred14} by 
removing the adversary's knowledge of side information, i.e., the non-sensitive attributes. 
Their idea is to poison the target model by injecting malicious data into the training set. 
By properly poisoning the target model, the model changes to the one that the adversary knows how to invert.

Yang et al. \cite{Yang19} proposed two main techniques for training inversion models in an adversarial setting. 
Unlike Fredrikson et al.'s optimization-based methods \cite{Fred14,Fred15}, 
Yang et al.'s method is training-based and inverts a model by learning a second model that acts as the inverse of the original one. 
The second model takes the predicted confidence score vectors of the original model as input, and outputs reconstructed data. 


In addition to the prevalent setting, model inversion attacks have also been investigated in other settings. 
Salem et al. \cite{Salem20} studied data reconstruction attacks in an online learning setting, 
where a target model is continuously updated using updating samples. 
By using the difference in the target model's outputs before and after an update, 
the updating samples can be reconstructed. 
Carlini et al. \cite{Carlini21} studied data extraction from language models. 
The attacker generates a large quantity of text data by sampling from the target model, 
and then launches a membership inference attack to select a text data point from the generated data as the reconstructed result.


\subsection{White-box model inversion attacks}

Zhang et al. \cite{Zhang20} presented a generative model inversion attack method 
that can invert DNNs and synthesize private training data with high fidelity. 
Their method involves two stages: public knowledge distillation and secret revelation. 
In the first stage, a generator and multiple discriminators 
are trained on public datasets to encourage the generator to create visually realistic images. 
In the second stage, the sensitive regions in the images created in the first stage 
are recovered by solving an optimization problem.

Geiping et al. \cite{Geiping20} investigated model inversion attacks in federated learning settings using users' gradients. 
The server separately stores and processes gradients transmitted by individual users. 
By using the norm magnitude and direction contained in gradients, 
the server can reconstruct high-quality data records. 
Yin et al. \cite{Yin21} improved Geiping et al's method \cite{Geiping20} by proposing a new method named GradInversion. 
GradInversion can recover a batch of images using average gradients. 
Specifically, they formulate the problem of input reconstruction from gradients as an optimization process, 
which converts random noise into natural images, matching gradients while regularizing image fidelity.

\subsection{Other related research}
In addition to the above experimental research, there is also theoretical research 
which formalizes model inversion attacks and investigates the factors that affect a model's vulnerability. 
Theoretical research focuses on discovering the underlying factors that affect the privacy of machine learning models 
but does not provide specific attack methods \cite{Wu16,Yeom18}. 

Other relevant research pertains to inversion from visual representations \cite{Mahen15,Doso16,Doso16CVPR}, 
where given the output of some intermediate layer of a neural network, a plausible input image is reconstructed. 
This type of research focuses on feature representations concentrating on properties of input images, 
while the research of model inversion centers on the privacy of training data and the properties of ML models.

There is also a stream of research that focuses on inferring the properties of training datasets \cite{Crist20,Ateniese15,Ganju18,Melis19}, 
such as the fraction of data that comes from a certain class. 
This stream of research typically pays attention to the global properties of training datasets 
instead of the privacy of individual data records that are targeted by model inversion attacks.

Moreover, reconstructing data in a model inversion attack is somewhat similar to the purpose of generative adversarial networks (GANs) 
\cite{Goodfellow14,Goodfellow16,Choi20}, as both seek to construct data. 
They do, however, have a significant difference. 
A typical GAN consists of a generator and a discriminator, 
where the generator creates examples while the discriminator tries to classify 
whether a given example is fake or real. 
In a GAN, the training data are not considered to be private information. 
Rather, they help the generator improve its quality. 
Whereas, with a model inversion attack, the training data are exactly what the model owner seeks to protect. 

\subsection{Summary of related work}
Existing model inversion attack methods must use information 
about either the output confidence of a model (black-box attacks) 
or the internal parameters of the model (white-box attacks). 
These methods, however, may not be efficient 
if a model releases only the label of an input data record. 
To overcome this limitation, we propose a novel attack method 
based only on predicted labels.

\section{Conclusion}\label{sec:conclusion}
This paper proposed a novel and effective inversion attack method based only on the labels of the input data records. 
Given an input data record, our method is conducted by training a shadow model to compute the highest score in the hidden confidence vector 
and then using the highest score to recover the hidden vector. 
Based on the recovered vector, our method trains an attack model to invert the input data record. 
Compared with related attack methods, our method enables an attacker to use minimal information to reconstruct input records with good quality. 
Our future work will focus on improving the inversion quality of input records by precisely recovering the confidence vectors. 
One possible approach is to incorporate generative techniques, e.g., GANs, to generate confidence vectors. 
Also, we will undertake research associated with defending against these model inversion attacks.


\bibliographystyle{plain}
{\small \bibliography{references}}

\appendix
\section{Additional experimental results}
This Appendix presents additional numerical results and the graphical results of the \emph{score-based} and \emph{one-hot} methods. 
\subsection{Effects of the prediction size}
\begin{figure}[ht]
\centering
	\includegraphics[scale=0.4]{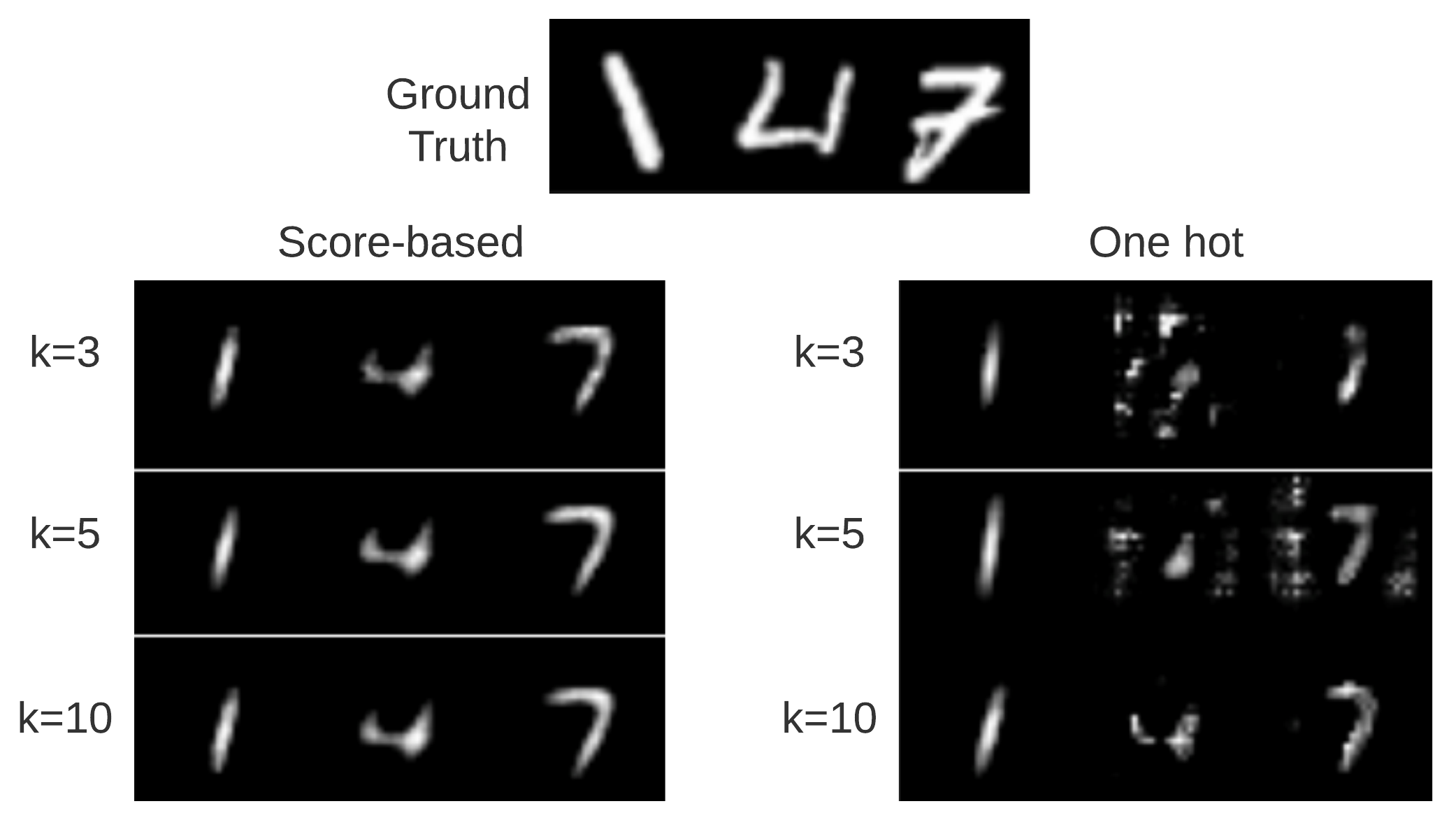}
	\caption{Performance of the \emph{score-based} and \emph{one-hot} methods on MNIST with different numbers of classes.
	Performance is evaluated using data records for $3$ classes, $5$ classes and $10$ classes.}
	\label{fig:MNIST-Class2}
\end{figure}

\begin{figure}[ht]
\centering
	\includegraphics[scale=0.4]{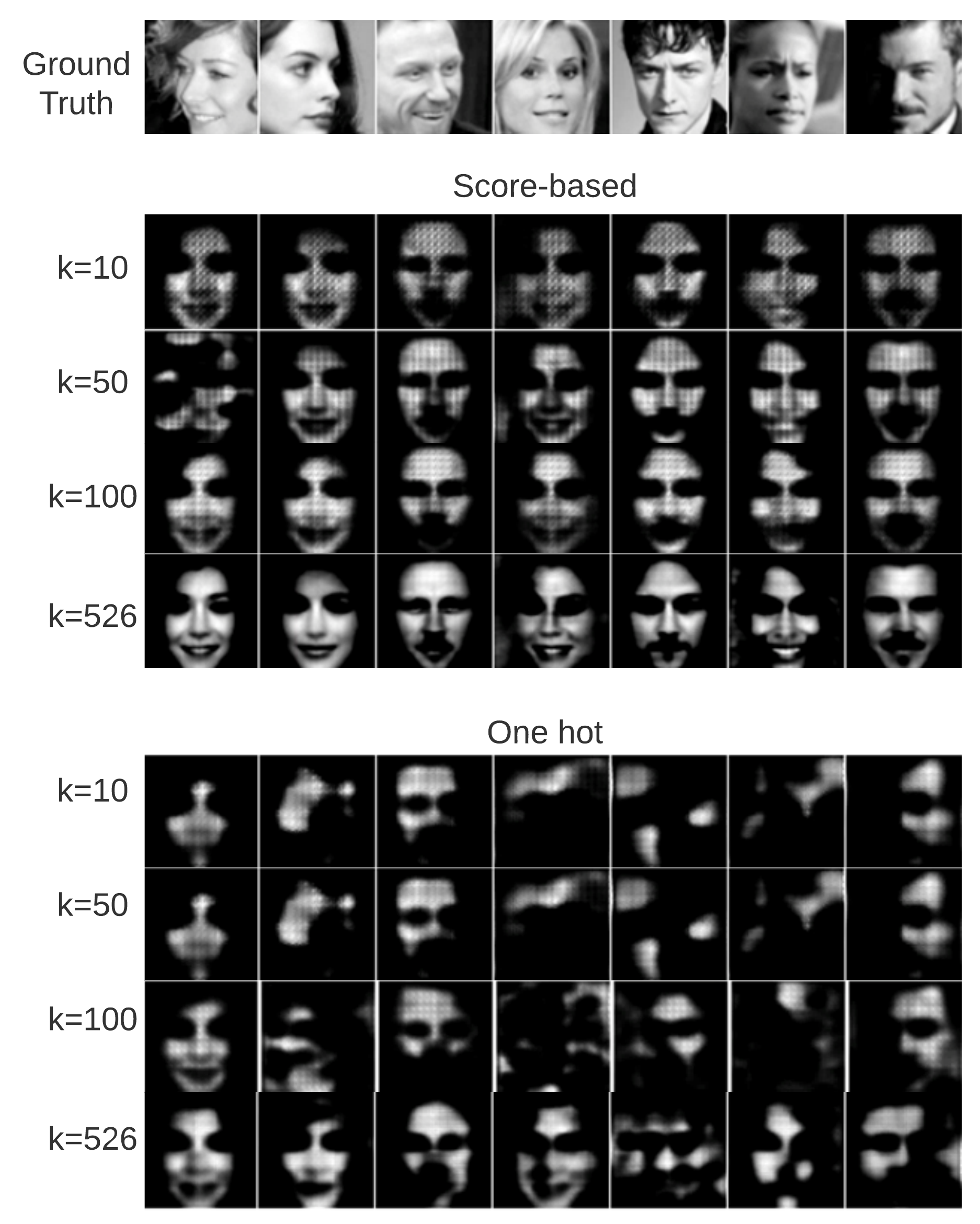}
	\caption{Performance of the \emph{score-based} and \emph{one-hot} methods on FaceScrub with different numbers of classes.
	Performance is evaluated using data records for $3$ classes, $5$ classes and $10$ classes.}
	\label{fig:FaceScrub-Class2}
\end{figure}

\begin{table}[!ht]\scriptsize
	\centering
	\caption{Data inversion errors of the four methods on MNIST and FaceScrub with different prediction size}
\begin{tabular}{|c|c|c|c|} \hline
& & MNIST & FaceScrub \\ \hline
\multirow{6}{*}{Label-only} & $3$ classes & $0.897206$ & ---- \\ \cline{2-4}
& $5$ classes & $0.897092$ & ---- \\ \cline{2-4}
& $10$ classes & $0.897076$ & $0.218111$ \\ \cline{2-4}
& $50$ classes & ---- & $0.209145$ \\ \cline{2-4}
& $100$ classes & ---- & $0.209009$ \\ \cline{2-4}
& $526$ classes & ---- & $0.208661$ \\ \hline
\multirow{6}{*}{Vector-based} & $3$ classes & $0.879813$ & ---- \\ \cline{2-4}
& $5$ classes & $0.873481$ & ---- \\ \cline{2-4}
& $10$ classes & $0.861832$ & $0.209487$ \\ \cline{2-4}
& $50$ classes & ---- & $0.187789$ \\ \cline{2-4}
& $100$ classes & ---- & $0.171219$ \\ \cline{2-4}
& $526$ classes & ---- & $0.170861$ \\ \hline
\multirow{6}{*}{Score-based} & $3$ classes & $0.886302$ & ---- \\ \cline{2-4}
& $5$ classes & $0.882318$ & ---- \\ \cline{2-4}
& $10$ classes & $0.880535$ & $0.215323$ \\ \cline{2-4}
& $50$ classes & ---- & $0.214261$ \\ \cline{2-4}
& $100$ classes & ---- & $0.213928$ \\ \cline{2-4}
& $526$ classes & ---- & $0.213561$ \\ \hline
\multirow{6}{*}{One-hot} & $3$ classes & $0.918335$ & ---- \\ \cline{2-4}
& $5$ classes & $0.912428$ & ---- \\ \cline{2-4}
& $10$ classes & $0.911258$ & $0.240116$ \\ \cline{2-4}
& $50$ classes & ---- & $0.238319$ \\ \cline{2-4}
& $100$ classes & ---- & $0.236435$ \\ \cline{2-4}
& $526$ classes & ---- & $0.232573$ \\ \hline
\end{tabular}
	\label{tab:class size}
\end{table}

\begin{table}[!ht]\scriptsize
	\centering
	\caption{Confidence vector recovery errors of the three methods on MNIST and FaceScrub with different numbers of classes to predict}
\begin{tabular}{|c|c|c|c|} \hline
& & MNIST & FaceScrub \\ \hline
\multirow{6}{*}{Label-only} & $3$ classes & $0.016$ & ---- \\ \cline{2-4}
& $5$ classes & $0.014$ & ---- \\ \cline{2-4}
& $10$ classes & $0.013$ & $0.056$ \\ \cline{2-4}
& $50$ classes & ---- & $0.055$ \\ \cline{2-4}
& $100$ classes & ---- & $0.052$ \\ \cline{2-4}
& $526$ classes & ---- & $0.051$ \\ \hline
\multirow{6}{*}{Score-based} & $3$ classes & $0.014$ & ---- \\ \cline{2-4}
& $5$ classes & $0.012$ & ---- \\ \cline{2-4}
& $10$ classes & $0.011$ & $0.050$ \\ \cline{2-4}
& $50$ classes & ---- & $0.049$ \\ \cline{2-4}
& $100$ classes & ---- & $0.046$ \\ \cline{2-4}
& $526$ classes & ---- & $0.045$ \\ \hline
\multirow{6}{*}{One-hot} & $3$ classes & $0.021$ & ---- \\ \cline{2-4}
& $5$ classes & $0.020$ & ---- \\ \cline{2-4}
& $10$ classes & $0.018$ & $0.069$ \\ \cline{2-4}
& $50$ classes & ---- & $0.068$ \\ \cline{2-4}
& $100$ classes & ---- & $0.067$ \\ \cline{2-4}
& $526$ classes & ---- & $0.065$ \\ \hline
\end{tabular}
	\label{tab:class size2}
\end{table}

This sub-section presents the results of \emph{score-based} and \emph{one-hot} methods on 
MNIST and FaceScrub with different numbers of classes to predict. 
As shown in Figures \ref{fig:MNIST-Class2} and \ref{fig:FaceScrub-Class2}, 
the performance of both the \emph{score-based} and \emph{one-hot} methods is independent of the prediction size to a large extent. 
This is because both only use one score while setting other scores to $0$. 
This result is also supported by the quantitative data inversion errors and confidence vector recovery errors 
as shown in Tables \ref{tab:class size} and \ref{tab:class size2}.  
We can see that the inversion errors and vector recovery errors of both the \emph{score-based} and \emph{one-hot} methods 
decrease slightly as the number of classes to predict increases. 

\subsection{Effects of the auxiliary set size}
\begin{figure}[ht]
\centering
	\includegraphics[scale=0.26]{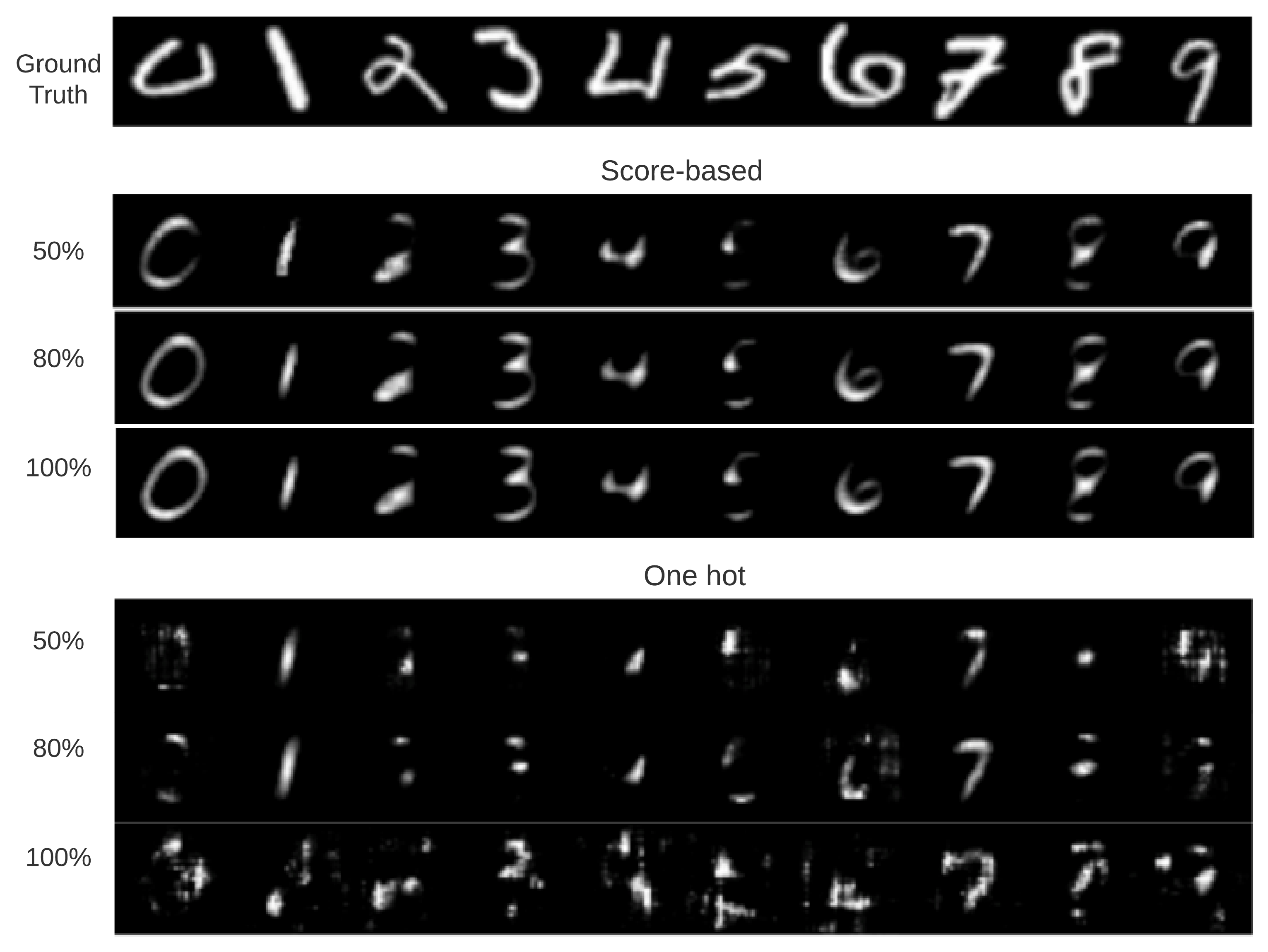}
	\caption{Performance of the \emph{score-based} and \emph{one-hot} methods on MNIST with different sized auxiliary sets. 
	Performance is evaluated using $50\%$, $80\%$ and $100\%$ of the auxiliary set.}
	\label{fig:MNIST-Set2}
\end{figure}

\begin{figure}[ht]
\centering
	\includegraphics[scale=0.26]{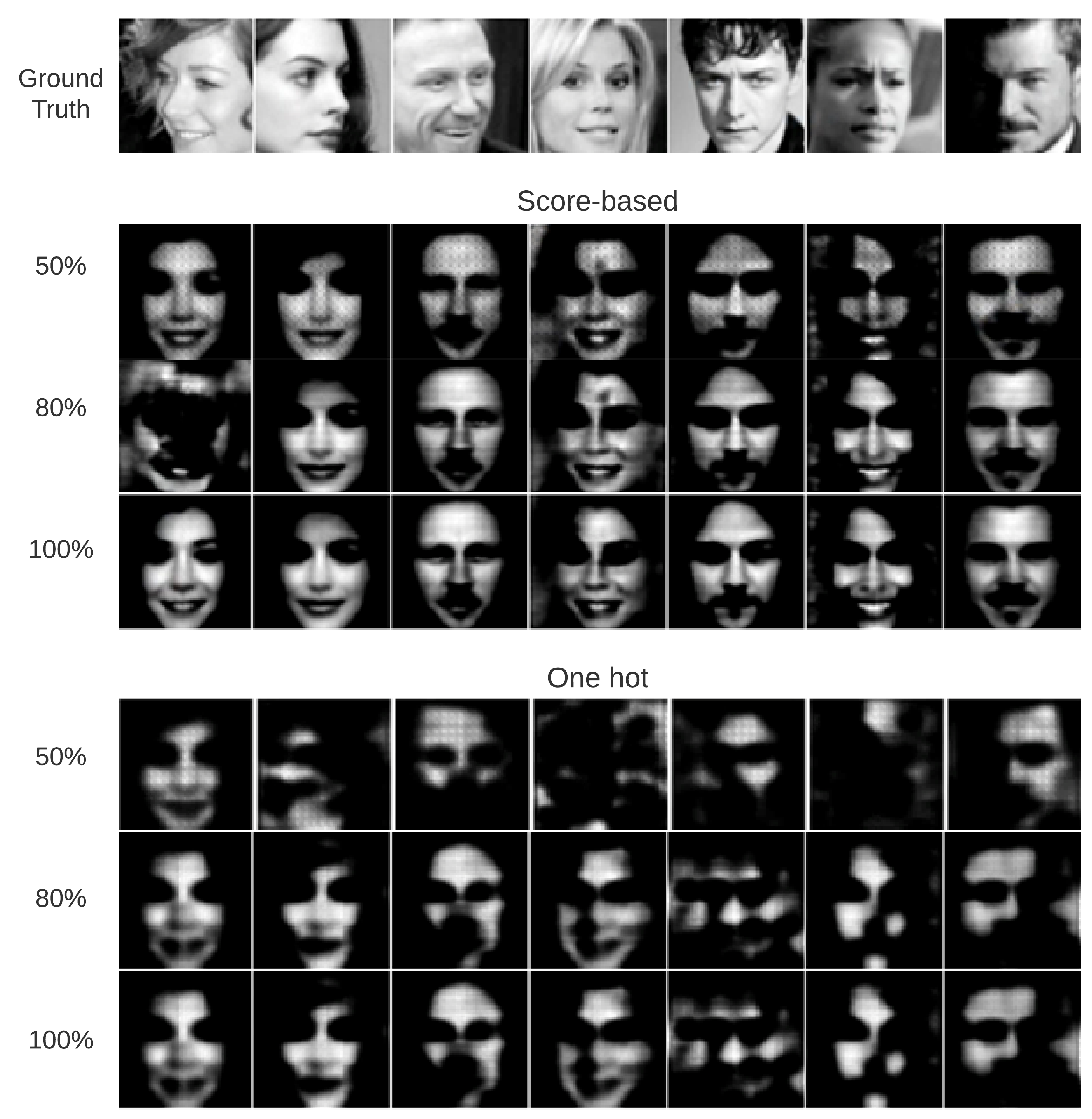}
	\caption{Performance of the \emph{score-based} and \emph{one-hot} methods on FaceScrub with different sized auxiliary sets. 
	Performance is evaluated using $50\%$, $80\%$ and $100\%$ of the auxiliary set.}
	\label{fig:FaceScrub-Set2}
\end{figure}

\begin{table}[!ht]\scriptsize
	\centering
	\caption{Data inversion errors of the four methods on MNIST and FaceScrub with different auxiliary set size}
\begin{tabular}{|c|l|l|l|} \hline
& & MNIST & FaceScrub \\ \hline
\multirow{3}{*}{Label-only} & $50\%$ auxil. set & $0.875275$ & $0.223795$ \\ \cline{2-4}
&$80\%$ auxil. set & $0.873224$ & $0.221577$ \\ \cline{2-4}
&$100\%$ auxil. set & $0.873628$ & $0.221443$ \\ \hline
\multirow{3}{*}{Vector-based} & $50\%$ auxil. set & $0.866639$ & $0.179698$ \\ \cline{2-4}
&$80\%$ auxil. set & $0.862412$ & $0.174564$ \\ \cline{2-4}
&$100\%$ auxil. set & $0.861832$ & $0.170806$ \\ \hline
\multirow{3}{*}{Score-based} & $50\%$ auxil. set & $0.873625$ & $0.218635$ \\ \cline{2-4}
&$80\%$ auxil. set & $0.870433$ & $0.217456$ \\ \cline{2-4}
&$100\%$ auxil. set & $0.868283$ & $0.217081$ \\ \hline
\multirow{3}{*}{One-hot} & $50\%$ auxil. set & $0.967396$ & $0.278733$ \\ \cline{2-4}
&$80\%$ auxil. set & $0.964241$ & $0.274545$ \\ \cline{2-4}
&$100\%$ auxil. set & $0.961328$ & $0.270437$ \\ \hline
\end{tabular}
	\label{tab:set size}
\end{table}

\begin{table}[!ht]\scriptsize
	\centering
	\caption{Confidence vector recovery errors of the three methods on MNIST and FaceScrub with different auxiliary set size}
\begin{tabular}{|c|l|l|l|} \hline
& & MNIST & FaceScrub \\ \hline
\multirow{3}{*}{Label-only} & $50\%$ auxil. set & $0.015$ & $0.054$ \\ \cline{2-4}
&$80\%$ auxil. set & $0.014$ & $0.053$ \\ \cline{2-4}
&$100\%$ auxil. set & $0.013$ & $0.051$ \\ \hline
\multirow{3}{*}{Score-based} & $50\%$ auxil. set & $0.014$ & $0.049$ \\ \cline{2-4}
&$80\%$ auxil. set & $0.012$ & $0.047$ \\ \cline{2-4}
&$100\%$ auxil. set & $0.011$ & $0.045$ \\ \hline
\multirow{3}{*}{One-hot} & $50\%$ auxil. set & $0.022$ & $0.067$ \\ \cline{2-4}
&$80\%$ auxil. set & $0.021$ & $0.065$ \\ \cline{2-4}
&$100\%$ auxil. set & $0.019$ & $0.064$ \\ \hline
\end{tabular}
	\label{tab:set size2}
\end{table}

This sub-section presents the results of \emph{score-based} and \emph{one-hot} methods on MNIST and FaceScrub with different auxiliary set sizes. 
As shown in Figures \ref{fig:MNIST-Set2} and \ref{fig:FaceScrub-Set2}, 
the performance of both the \emph{score-based} and \emph{one-hot} methods is independent of the auxiliary set size to a large extent,  
as both of them aim to recover only generic features of images. 
This result is also supported by the quantitative data inversion errors and confidence vector recovery errors 
as shown in Tables \ref{tab:class size} and \ref{tab:class size2},   
where the inversion errors and vector recovery errors of both the \emph{score-based} and \emph{one-hot} methods 
decrease slightly as the size of the auxiliary set increases. 

\subsection{Effects of the auxiliary set distribution}
\begin{figure}[ht]
\centering
	\includegraphics[scale=0.43]{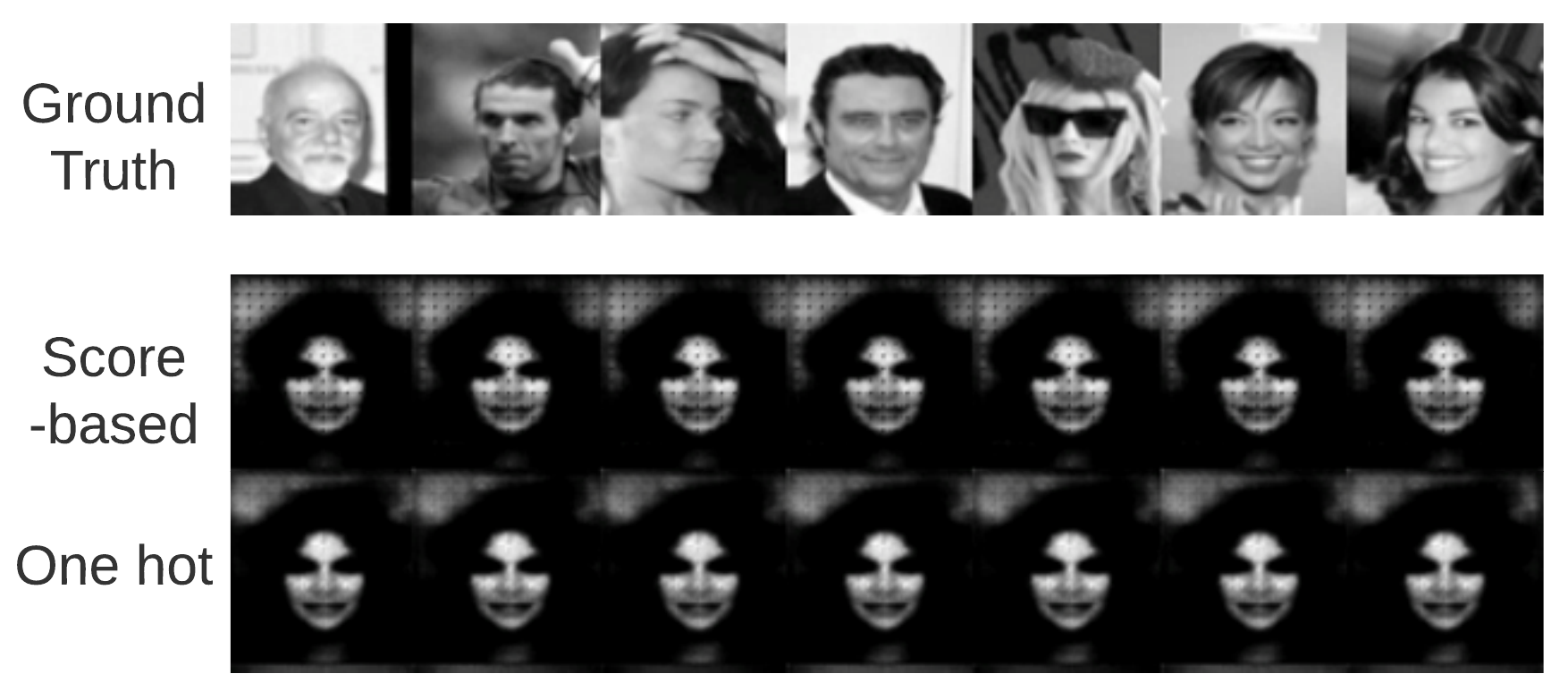}
	\caption{Performance of the \emph{score-based} and \emph{one-hot} methods when using FaceScrub to train the target model and CelebA to train attack models.}
	\label{fig:FaceScrub-distribution2}
\end{figure}
In Figure \ref{fig:FaceScrub-distribution2}, akin to our \emph{label-only} method, 
both the \emph{score-based} and \emph{one-hot} methods reconstruct only ``average'' faces. 
This is because the target model has never seen the classes in CelebA.
Hence, the target model's outputs, with respect to CelebA samples, are meaningless. 
Using these outputs to train the attack models results in an incorrect model. 

\subsection{Effects of attack model structure}
\begin{figure}[ht]
\centering
	\includegraphics[scale=0.38]{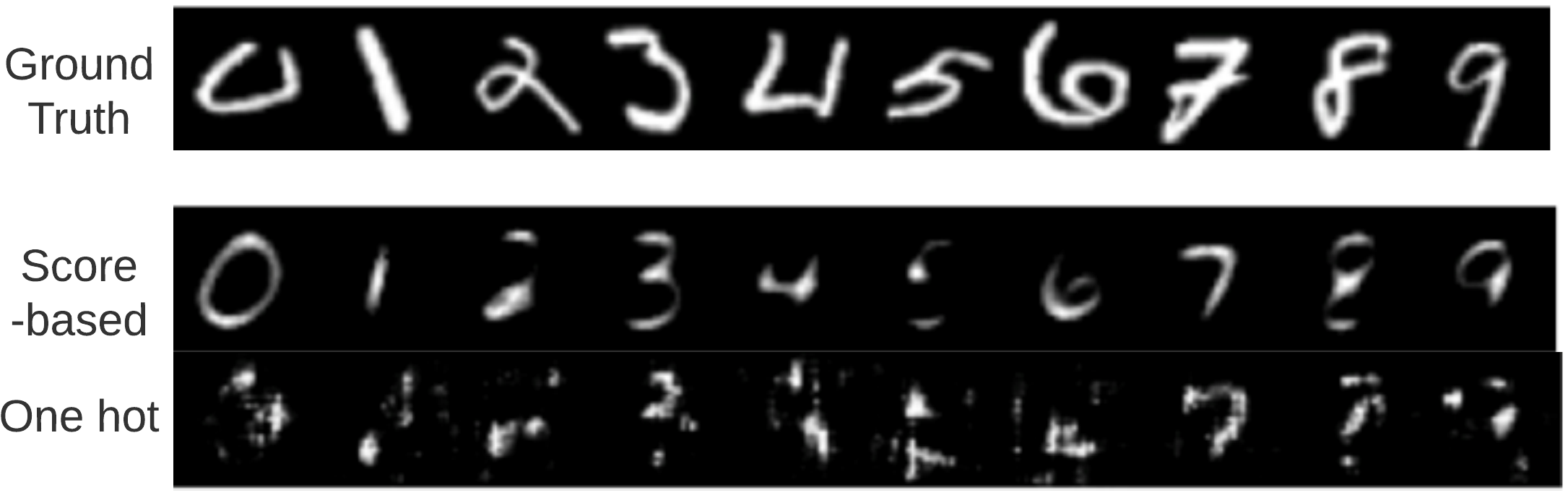}
	\caption{Performance of the \emph{score-based} and \emph{one-hot} methods on MNIST when removing a transposed CNN block from the attack models.}
	\label{fig:MNIST-structure2}
\end{figure}

\begin{figure}[ht]
\centering
	\includegraphics[scale=0.43]{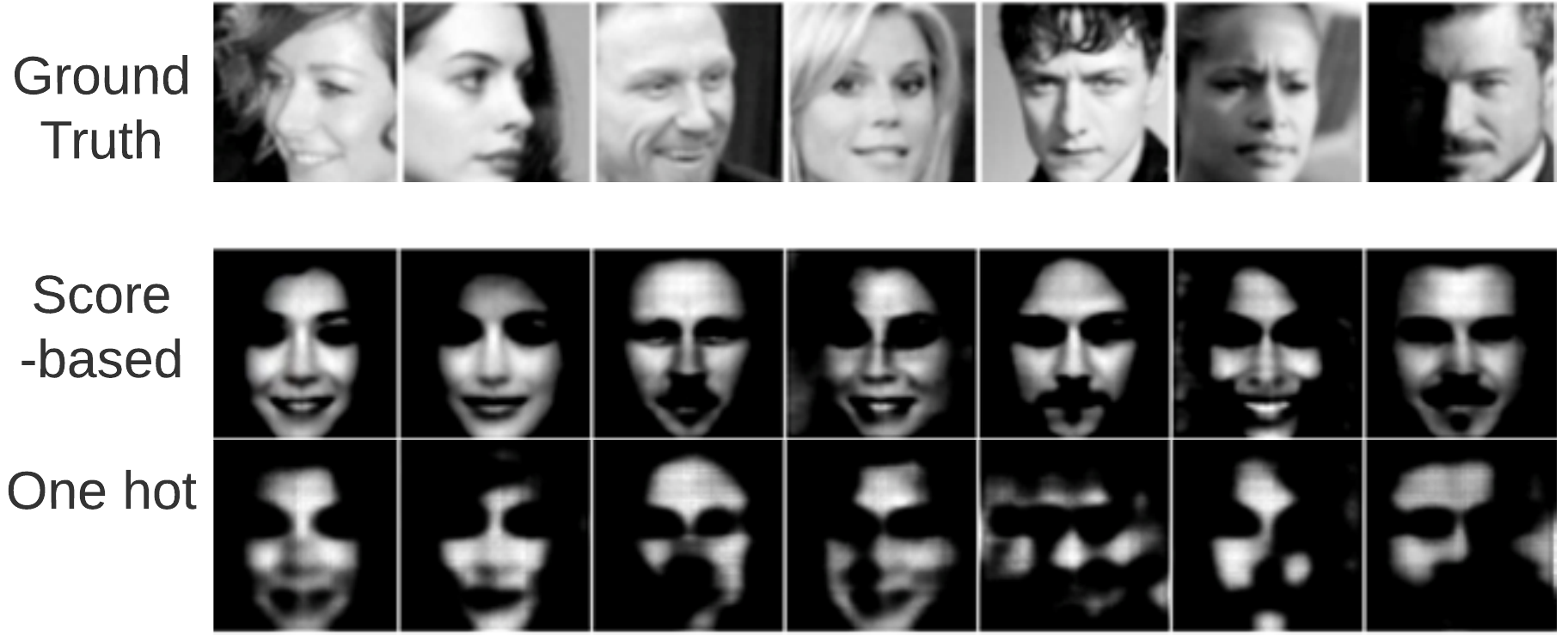}
	\caption{Performance of the \emph{score-based} and \emph{one-hot} methods on FaceScrub when removing a transposed CNN block from the attack models.}
	\label{fig:FaceScrub-structure2}
\end{figure}

Figures \ref{fig:MNIST-structure2} and \ref{fig:FaceScrub-structure2} demonstrate the inversion results of 
the \emph{score-based} and \emph{one-hot} methods on MNIST and FaceScrub, respectively, when a transposed CNN block from attack models is removed. 
We can see that such a modification of attack models does not have much impact on the performance of both methods. 
This is due to the aim of the two methods, 
where the structures of attack models are not very sensitive to recovering the generic features of images.

\end{document}